\documentclass[prd,aps,
floats,letterpaper,floatfix,
groupedaddress,superscriptaddress
,nofootinbib
,eqsecnum
,twocolumn
]{revtex4}
\usepackage{amsmath}
\usepackage{amsfonts}
\usepackage{bm}
\usepackage{graphicx}
\usepackage[usenames]{color}
\newcommand{\gothg}{\mathfrak{g}}
  \newcommand{\apjl}{Astrophys.\ J.\ Lett.}

 \newcommand{\jcap}{J.\ Cosmol.\ Astropart.\ Phys.}
 \newcommand{\cqg}{Class.\ Quantum Grav.}
 \newcommand{\lrr}{Living Rev.\ Relativ.}

\allowdisplaybreaks
\def\eae{Einstein-{\AE}ther }

\begin{document}

\title{Compact binary systems in Einstein-{\AE}ther gravity: Direct integration of the relaxed field equations to 2.5 post-Newtonian order}
\author{
Fatemeh Taherasghari} \email{ftaherasghari@ufl.edu}
\affiliation{Department of Physics, University of Florida, Gainesville, Florida 32611, USA}

\author{
Clifford M.~Will} \email{cmw@phys.ufl.edu}
\affiliation{Department of Physics, University of Florida, Gainesville, Florida 32611, USA}
\affiliation{GReCO, Institut d'Astrophysique de Paris, CNRS,\\ 
Sorbonne Universit\'e, 98 bis Bd.\ Arago, 75014 Paris, France}

\date{\today}

\begin{abstract}
The \eae theory is an alternative theory of gravity in which the spacetime metric is supplemented by a long-range timelike vector field (the ``aether'' field). 
Here, for the first time, we apply the full formalism of post-Minkowskian theory and of  the Direct Integration of the Relaxed Einstein Equations (DIRE), to this theory of gravity, with the goal of deriving equations of motion and gravitational waveforms for orbiting compact bodies to high orders in a post-Newtonian expansion.   Because the aether field is constrained to have unit norm, a naive application of post-Minkowskian theory leads to contributions to the effective energy momentum tensor that are {\em linear} in the perturbative fields.  We show that a suitable redefinition of fields using an array of ``superpotentials'' can eliminate such linear terms to any desired post-Newtonian order, resulting in flat spacetime wave equations for all fields, with sources consisting of matter terms and terms quadratic and higher in the fields.  As an initial application of this new method, and as a foundation for obtaining the equations of motion for compact binaries, we obtain explicit solutions of the relaxed equations sufficient to obtain the metric in the near zone through 2.5 post-Newtonian order, or $O[(v/c)^5]$ beyond the Newtonian approximation.  

\end{abstract}

\maketitle

\section{Introduction}
\label{sec:intro}

One of the classic approaches to devising a theory of gravity alternative to general relativity (GR) is to postulate, in addition to the spacetime metric, an auxiliary gravitational field.  The quintessential example is the 1961 Brans-Dicke theory (which built upon earlier work by Fierz, Pauli and Jordan)\cite{1961PhRv..124..925B}, in which the added field was a scalar.  By proposing a suitable action for the auxiliary field along with a suitable coupling between it and the action for the spacetime metric, one could obtain field equations with reasonable mathematical properties (such as partial differential equations of order no greater than two).  In addition, one could automatically abide by very precise tests of the Einstein Equivalence Principle, such as the E\"otv\"os experiment, by ensuring that the coupling to the fields of matter involved only the spacetime metric, a concept called ``universal coupling'' or ``metric coupling''.  This set of ideas continues to serve as a template for inventing theories of gravity into the present, with a profusion of theories having multiple scalar fields, vector fields, and tensor fields of various ranks  (for reviews, see 
\cite{2007sttg.book.....F,
2010RvMP...82..451S,
2010LRR....13....3D,
2014LRR....17....7D,
2014LRR....17....4W,
2015CQGra..32x3001B,
2018LRR....21....1B}).

One of the earliest {\em vector-tensor} theories was invented by Will and Nordtvedt 
\cite{1972ApJ...177..757W} and later generalized by Hellings and Nordtvedt \cite{1973PhRvD...7.3593H}, motivated by a desire to explore theories that might exhibit ``preferred-frame'' effects.  In general relativity, the gravitational physics of an isolated system does not depend on its velocity relative to the rest of the universe because the asymptotic, or large-distance limit of the metric (which establishes the boundary conditions for solving for the local gravitational physics) can always be transformed to the Minkowski metric, which is independent of the motion of the reference frame in which it is observed.  The same is true in scalar-tensor theories because the asymptotic scalar field is also independent of reference frame.  By contrast, in a theory with a timelike vector field $K^\mu$ that is somehow related to the distribution of mass energy, the asymptotic field that establishes the boundary conditions for that system would be expected to point purely in the time direction (i.e. have components $(K^0,\, 0,\,0,\,0)$) if the system is at rest relative to the mean rest frame of the cosmic distribution of matter.  But if an isolated system were to move relative to that cosmic frame, then the asymptotic vector field in the frame of the system would have the form $(K^0,\, K^1,\, K^2,\, K^3)$, where the spatial part of the vector field is related to the speed and direction of motion relative to the cosmic frame  (see Chapter 5 of \cite{tegp2} for a review of alternative theories of gravity), and this would alter the internal structure and dynamics of the isolated system.  

One defect of these early theories was that the field equation for the vector field was homogeneous and linear in $K^\mu$ with no matter source (by virtue of metric coupling), so that $K^\mu = 0$ was an immediate solution unless one forced the asymptotic value of $K^0$ or $|K^\mu|$ to be a non-zero arbitrary constant.

As a result, the subject of vector-tensor theories lay somewhat dormant until Jacobson and colleagues proposed the ``\eae'' theory \cite{2001PhRvD..64b4028J,2002cls..conf..331M,2004PhRvD..70b4003J,2004PhRvD..69f4005E,2006PhRvD..73f4015F}.  As before, the goal was to study
violations of Lorentz invariance in gravity, now in parallel
with similar studies in matter interactions, such as the Standard Model Extension of Kostaleck\'y and Samuel 
 \cite{1989PhRvD..40.1886K}.  Another motivation was the notion that such Lorentz violations might be a classical relic of a quantum gravity theory in which there was a fundamental quantum of length.   Other theories and generalizations followed, including the Tensor-Vector-Scalar (TeVeS) theory of Bekenstein \cite{2004PhRvD..70h3509B}, designed to provide a relativistic foundation for the phenomenological Modified Newtonian Dynamics (MOND) proposal of Milgrom\cite{1983ApJ...270..365M}; Khronometric theory, a low-energy limit of ``Ho\v{r}ava gravity'', a proposal for a theory that is power-counting renormalizable \cite{2009PhRvD..79h4008H}, shown later to be a singular limit of \eae theory \cite{2010PhRvL.104r1302B,2011JHEP...04..018B,2014PhRvD..89h1501J}; the Scalar-Tensor-Vector (STV, but also called MOG) of Moffat \cite{2006JCAP...03..004M}, designed to avoid the need for dark matter; and a generalized tensor-vector-scalar theory of Skordis \cite{2008PhRvD..77l3502S,2009CQGra..26n3001S}, designed mainly for cosmological investigations.

At the lowest post-Newtonian (PN) order, the parametrized post-Newtonian (PPN) parameters of \eae theory were calculated by Foster and Jacobson \cite{2006PhRvD..73f4015F}; the values were identical to those of general relativity, except for the ``preferred-frame'' parameters, $\alpha_1$ and $\alpha_2$, which could be non-zero.  Foster also derived the leading gravitational radiation damping effects  \cite{2006PhRvD..73j4012F,2007PhRvD..75l9904F} and the PN equations of motion for compact bodies such as neutron stars and black holes \cite{2007PhRvD..76h4033F}, later verified by Yagi et al.\ \cite{2014PhRvD..89h4067Y}.   Constraints on the parameters of the theory have been placed using binary pulsar data
\cite{2014PhRvL.112p1101Y,2014PhRvD..89h4067Y,2021CQGra..38s5003G}.

The detection of gravitational waves from inspiralling binary black holes in 2015 presented new possibilities for testing alternative theories of gravity, and the LIGO-Virgo collaboration has published comprehensive papers detailing a wide range of tests, first using data from the discovery event GW 150914 \cite{2016PhRvL.116v1101A}, and subsequently using data from the full catalogue of events through the middle of the third observing run \cite{2021PhRvD.103l2002A}.  
One notable result was the observation of the nearly concident arrival times of the gravitational-wave and gamma-ray signals from the binary neutron star merger event GW170817/GRB170817 \cite{2017PhRvL.119p1101A,2017ApJ...848L..13A}, which placed an extremely strong bound on the speed of gravitational waves, relative to that of light,
\begin{equation}
- 3 \times 10^{-15} < v_g -1 < 7 \times 10^{-16} \,.
\end{equation}
This had the effect of ruling out a significant number of alternative theories of gravity
\cite{2017PhRvL.119y1301B,2017PhRvL.119y1302C,2017PhRvL.119y1304E,2017PhRvL.119y1303S}, and constraining the Einstein-{\AE}ther Theory
\cite{2018PhRvD..97l4023O}.  

Gravitational-wave data have also constrained the strong-field dynamical evolution of compact binary mergers, as reflected in the detailed time evolution of the detected waveforms.  No deviations from the predictions of general relativity have been found, and constraints have been placed on the coefficients of the terms in a PN expansion of the waveform phase \cite{2021PhRvD.103l2002A}.  While these ``theory agnostic'' constraints are useful and important, they provide only limited information about what theories might be ruled out, simply because very few theories have been analyzed in sufficient detail to provide predictions for these coefficients to an order comparable to what is known for GR.  In scalar-tensor theories, considerable effort has gone toward obtaining the coefficents up to 2PN order 
\cite{2013PhRvD..87h4070M,2014PhRvD..89h4014L,2015PhRvD..91h4027L,2016PhRvD..94h4003S,2018PhRvD..98d4004B,2022JCAP...08..008B}.  However, because of the very strong bound on the scalar-tensor coupling parameter $\omega$ from solar-system measurements, combined with the fact that, in this class of theories, binary-black hole evolution is indistinguishable from its counterpart in general relativity, it seems unlikely that gravitational-wave measurements will lead to stronger constraints, except possibly via the detection of a favorable black-hole neutron-star merger.  

What makes the study of gravitational waves in alternative theories intriguing is that they generally predict the existence of {\em dipole} and even {\em monopole} gravitational radiation, none of which exist in GR.  In particular, if the binary source is sufficently asymmetrical, either in mass or composition (eg.\ a black-hole neutron-star binary), then dipole gravitational radiation can lead to contributions to the energy flux and the waveform evolution that are larger than the conventional quadrupole contributions by a factor of $(c/v)^2$, where $v$ is the orbital velocity.  In other words, dipole radiation effects can occur at ``-1PN'' order, in a hierarchy where quadrupole radiation is denoted by ``0PN'' order.  This is both a blessing and a curse.  It is a blessing because it could lead to tighter constraints on the theory than might have been expected a priori.  But it is a curse because, in order to calculate the waveform evolution to an order equivalent to the $n$PN order of general relativity, one must determine the radiative moments of the auxiliary fields and the equations of motion of the binary system to the $(n+1)$PN order.  

These considerations have motivated us to begin an effort to determine the equations of motion for compact binaries and the emitted gravitational-waveform in a post-Newtonian expansion of Einstein-{\AE}ther theory beyond the lowest-order dipole and quadrupole contributions, and beyond linearized theory, which constitute the current state of the art \cite{2020PhRvD.101d4002Z,2023arXiv230406801S}.  Because of the significant additional complexity of this class of theories, combined with the ``curse'' of dipole radiation, our goal will be modest: to obtain the gravitational waveform to 1.5 PN order beyond the conventional quadrupole level.  This paper is devoted to obtaining the metric to 2.5PN order, while future papers will obtain the equations of motion for compact bodies to 2.5PN order, the far-zone fields to the required order, and finally the energy flux and waveform to 1.5PN order.  The results will augment the waveform templates described in 
\cite{2020PhRvD.101d4002Z,2023arXiv230406801S}.

In Sec.\ \ref{sec:EAE theory}, we review the essentials of Einstein-{\AE}ther theory, and impose a condition on one of its four arbitrary parameters that arises from the gravitational-wave speed constraint from GW170817.  Section \ref{sec:relaxed} expresses the theory in the form of ``relaxed field equations'' of the post-Minkowskian method that has been used in GR and scalar-tensor theory to carry out PN expansions (see eg. \cite{PW2014}).  In Sec.\ \ref{sec:nearzonefields} we note that the presence in Einstein-{\AE}ther theory of a vector auxiliary field with unit norm  necessitates a change of field variables in order to obtain wave equations for the fields whose sources consists of matter plus field contributions that are quadratic in small quantities, thus enabling a consistent PN expansion.  Section \ref{sec:2.5expansion} obtains solutions within the near-zone for the fields to orders that permit the construction of the complete spacetime metric to 2.5PN order.  In Sec.\ \ref{sec:conclusions} we briefly describe ongoing work and make concluding remarks. 

\section{Einstein-{\AE}ther Theory}
\label{sec:EAE theory}

\eae  theory is defined by the covariant action
\begin{align}
S_{\text{\AE}} &\equiv \frac{1}{16\pi G_0} \int \sqrt{-g} \biggl [ R -E^{\mu\nu}_{\alpha\beta} \nabla_\mu K_{\rm ae}^\alpha \nabla_\nu K_{\rm ae}^\beta 
\nonumber \\ 
&\quad \quad + \lambda(K_{{\rm ae} \,\mu} K_{\rm ae}^\mu +1) \biggr ] d^4x 
+ \int \sqrt{-g} \,{\cal L}_M d^4x \,,
\label{eq:aeaction}
\end{align}
where $g$ is the determinant of the metric $g_{\mu\nu}$, $R$ is the Ricci scalar, $\nabla_\mu$ is a covariant derivative with respect to the metric,
\begin{equation}
E^{\mu\nu}_{\alpha\beta} = c_1 g^{\mu\nu}g_{\alpha\beta} + c_2 \delta^\mu_\alpha \delta^\nu_\beta + c_3  \delta^\mu_\beta \delta^\nu_\alpha  - c_4 K_{\rm ae}^\mu K_{\rm ae}^\nu g_{\alpha\beta} \,,
\end{equation}
$\lambda$ is a Lagrange multiplier designed to enforce the constraint $K_{{\rm ae}\,\mu} K_{\rm ae}^\mu = -1$, and ${\cal L}_M$ is the matter Lagrangian.  We use units in which the speed of light $c$ is unity, and the spacetime metric has the signature $(-,+,+,+)$; Greek indices denote spacetime components and Roman indices denote spatial components; parentheses (square brackets) around groups of indices denote symmetrization (antisymmetrization).  

However, it is well-known that the speed of transverse-traceless gravitational waves in this theory is given by $v_g = (1-c_1-c_3)^{-1/2}$.  Because of the extraordinary bound from GW170817, {\em we will make the assumption that} $c_3 = -c_1$, reducing the theory to a three-parameter set of theories.  The Lagrangian for the {\AE}ther field then takes the form
\begin{equation}
{\cal L}_{\text{\AE}} = -\frac{1}{2} c_1 F_{\mu\nu} F^{\mu\nu} -c_2 \left (\nabla_\mu K_{\rm ae}^\mu \right )^2 +c_4 | \bm{a}_{\rm ae} |^2  \,,
\end{equation}
where 
\begin{equation}
F_{\mu\nu} \equiv \partial_\nu K_{{\rm ae}\,\mu} -  \partial_\mu K_{{\rm ae}\,\nu} \,,
\qquad
a_{\rm ae}^\mu \equiv K_{\rm ae}^\nu \nabla_\nu K_{\rm ae}^\mu \,.
\end{equation}

The initial studies by Jacobson et al.\ established the post-Newtonian limit, studied gravitational wave propagation, and analyzed other aspects of the theory.  Here we summarize the main results, {\em but we impose the constraint $c_3 = -c_1$ a priori}.
The parametrized post-Newtonian (PPN) parameters \cite{tegp2} are given by \cite{2006PhRvD..73f4015F}:
\begin{align}
  \gamma & =  1 \,, \quad  \beta  =  1\,,
\nonumber \\
  \alpha_1 & =  - 4 c_{14} \,,
\nonumber  \\
  \alpha_2 & =  \frac{c_{14}}{c_2} \left (\frac{2c_2c_{14}+c_{14} -c_2}{2-c_{14}} \right )\,,
   \label{eq5:PPNAE}
\end{align}
with $ \xi  =  \alpha_3 = \zeta_1 = \zeta_2 = \zeta_3 = \zeta_4 = 0$,
where $c_{14}=c_1+c_4$.
The present value of the gravitational constant is given by
\begin{equation}
G = G_0 \left (1-\frac{c_{14}}{2} \right )^{-1}  \,.
\label{eq5:GAEtheory}
\end{equation}
Considerations of positivity of energy impose the constraints $c_1 \ge 0$, $c_2 \ge 0$,
and $c_{14} \ge 0$.

\section{The relaxed field equations in \eae theory}
\label{sec:relaxed}

\subsection{Field equations}
\label{sec:fieldequations}

We begin by deriving the field equations in a form that will be useful for obtaining the so-called ``relaxed'' field equations, analogous to those in general relativity.
Varying the action (\ref{eq:aeaction})  with respect to the  \AE ther field yields the field equation for $K_{\rm ae}^\mu$,
\begin{align}
c_1 \nabla_\nu F^{\mu\nu} &= 8\pi G_0 T_{\rm ae}^\mu 
-\lambda K_{\rm ae}^\mu- c_2 \nabla^\mu (\nabla_\nu K_{\rm ae}^\nu)
\nonumber \\
& \quad
-c_4 \left ( a_{\rm ae}^\nu \nabla^\mu K_{{\rm ae}\,\nu} -  a_{\rm ae}^\mu \nabla_\nu K_{\rm ae}^\nu - K_{\rm ae}^\nu \nabla_\nu  a_{\rm ae}^\mu \right )
 \,,
\label{eq:aether1}
\end{align}
where we define the matter energy-momentum tensor and vector by
\begin{equation}
T^{\mu\nu}  \equiv  \frac{2}{\sqrt{-g}} \frac{\delta (\sqrt{-g} {\cal L}_M )}{\delta g_{\mu\nu} } \,, \quad
{T_{\rm ae}}_\mu \equiv - \frac{1}{\sqrt{-g}} \frac{\delta (\sqrt{-g} {\cal L}_M )}{\delta K_{\rm ae}^\mu} \,.
\end{equation}
In a conventional metric theory of gravity, where the matter Lagrangian couples only to the metric, the quantity ${T_{\rm ae}}_\mu$ would vanish.  However we anticipate treating compact bodies, where the mass of each body may have an {\em effective} dependence on $K_{\rm ae}^\mu$ via its gravitational binding energy.  This idea is based on the original proposal by Eardley \cite{1975ApJ...196L..59E}, with follow-up work by Gralla \cite{2010PhRvD..81h4060G,2013PhRvD..87j4020G}; see also \cite{2022PhRvD.106f4021T}.  This will be addressed in detail in a subsequent paper; for now we will include the effective energy-momentum vector in all our considerations.

Contracting Eq.\ (\ref{eq:aether1}) with $K_{{\rm ae}\,\mu}$ and using the constraint that $K_{{\rm ae}\,\mu} K_{\rm ae}^\mu = -1$ yields an expression for the Lagrange multiplier $\lambda$
\begin{align}
\lambda &= -8 \pi G_0 K_{{\rm ae}\,\nu} T_{\rm ae}^\nu - \frac{1}{2} c_1 \left ( F_{\mu\nu} F^{\mu\nu} +2 \nabla_\nu a_{\rm ae}^\nu \right )
\nonumber \\
& \quad \quad +c_2 K_{\rm ae}^\nu \nabla_\nu (\nabla_\mu K_{\rm ae}^\mu) + 2c_4 |{\bm a}_{\rm ae} |^2 \,.
\label{eq:lambda}
\end{align}
Note that ${\bm K}_{\rm ae} \cdot {\bm a}_{\rm ae} =0$.
Varying the action with respect to the metric and making use of Eq.\ (\ref{eq:lambda}) to eliminate the Lagrange multiplier $\lambda$, we obtain the field equations
\begin{equation}
G^{\mu\nu} = 8 \pi G_0 \left ( T^{\mu\nu}  - K_{\rm ae}^\mu K_{\rm ae}^\nu K_{{\rm ae}\,\alpha} T_{\rm ae}^\alpha \right ) + S^{\mu\nu} \,,
\label{eq:fieldeq}
\end{equation}
for the metric, and
\begin{align}
c_1 \nabla_\nu F^{\mu\nu} &= 8\pi G_0 \left (T_{\rm ae}^\mu+ K_{\rm ae}^\mu K_{{\rm ae}\,\nu} T_{\rm ae}^\nu \right )
\nonumber \\
&+ \frac{1}{2} c_1 K_{\rm ae}^\mu \left ( F_{\alpha\beta} F^{\alpha\beta}  +2 \nabla_\nu a_{\rm ae}^\nu \right )
\nonumber \\
&- c_2 \left [ \nabla^\mu (\nabla_\nu K_{\rm ae}^\nu) +K_{\rm ae}^\mu K_{\rm ae}^\nu \nabla_\nu (\nabla_\alpha K_{\rm ae}^\alpha) \right ]
\nonumber \\
& 
-c_4 \left [ a_{\rm ae}^\nu \nabla^\mu K_{{\rm ae}\,\nu} -  a_{\rm ae}^\mu \nabla_\nu K_{\rm ae}^\nu 
\right .
\nonumber \\
& \left . \qquad
- K_{\rm ae}^\nu \nabla_\nu  a_{\rm ae}^\mu +2 K_{\rm ae}^\mu |{\bm a}_{\rm ae} |^2 \right ] \,,
\label{eq:aether2}
\end{align}
for the {\AE}ther field, where
\begin{align}
S^{\mu\nu} &= c_1 \biggl [ 2K_{\rm ae}^{(\mu} \nabla_\alpha F^{\nu ) \alpha}  +{ F_\alpha}^\mu F^{\alpha\nu} 
-K_{\rm ae}^\mu K_{\rm ae}^\nu (\nabla_\alpha a_{\rm ae}^\alpha )
\nonumber \\
& \qquad
- \frac{1}{4} \left ( g^{\mu\nu} + 2 K_{\rm ae}^\mu K_{\rm ae}^\nu \right )F_{\alpha\beta} F^{\alpha\beta} 
 \biggr ]
\nonumber \\
& 
\quad +c_2 \biggl [ 
     \left (g^{\mu\nu}+K_{\rm ae}^\mu K_{\rm ae}^\nu \right )K_{\rm ae}^\beta \nabla_\beta (\nabla_\alpha K_{\rm ae}^\alpha)
\nonumber \\
&
\qquad 
+\frac{1}{2} g^{\mu\nu} (\nabla_\alpha K_{\rm ae}^\alpha )^2 
     \biggr ]
\nonumber \\
& \quad
     -c_4 \biggl [ a_{\rm ae}^\mu a_{\rm ae}^\nu - \frac{1}{2} \left (g^{\mu\nu}+4 K_{\rm ae}^\mu K_{\rm ae}^\nu \right ) |{\bm a}_{\rm ae} |^2 
\nonumber \\
& \qquad
    -K_{\rm ae}^\mu K_{\rm ae}^\nu ( \nabla_\alpha a_{\rm ae}^\alpha ) 
    - 2 a_{\rm ae}^\alpha (\nabla_\alpha K_{\rm ae}^{(\mu} ) K_{\rm ae}^{\nu )}
\nonumber \\
& \qquad        
    +2 a_{\rm ae}^{(\mu} K_{\rm ae}^{\nu )} \nabla_\alpha K_{\rm ae}^\alpha  + 2K_{\rm ae}^\alpha (\nabla_\alpha a_{\rm ae}^{(\mu} ) K_{\rm ae}^{\nu )}  \biggr ] \,.
 \end{align}
  Note that contracting Eq.\ (\ref{eq:aether2}) with $K_{{\rm ae}\,\mu}$ now yields a trivial equality.

\subsection{Relaxed \eae field equations}
\label{sec:relaxed2}

To recast Eq.~(\ref{eq:fieldeq}) into the form of a ``relaxed'' \eae equation, we define the 
quantities 
\begin{eqnarray}
\gothg^{\mu\nu} &\equiv& \sqrt{-g} g^{\mu\nu} \,, 
\nonumber\\
H^{\mu\alpha\nu\beta} &\equiv& \gothg^{\mu\nu}  \gothg^{\alpha\beta} -  \gothg^{\alpha\nu} \gothg^{\beta\mu} \,,
\label{Hdef}
\end{eqnarray}
and use the identity, valid for any spacetime,
\begin{equation}
{H^{\mu\alpha\nu\beta}}_{,\alpha\beta} = (-g) (2G^{\mu\nu} + 16\pi t_{LL}^{\mu\nu} ) \,,
\label{Hidentity}
\end{equation}
where $t_{LL}^{\mu\nu}$ is the Landau-Lifshitz pseudotensor.
We next define the gravitational field ${h}^{\mu\nu}$ by the equation
\begin{equation}
{\gothg}^{\mu\nu} \equiv  \eta^{\mu\nu} - {h}^{\mu\nu} \,,
\label{hdef}
\end{equation}
and impose the ``Lorenz''  or harmonic gauge condition
\begin{equation}
{{h}^{\mu\nu}}_{,\nu} = 0 \,.
\label{gauge}
\end{equation}
Substituting Eqs.\ (\ref{eq:fieldeq}), (\ref{hdef}) and (\ref{gauge}) into (\ref{Hidentity}), we can recast the field equation (\ref{eq:fieldeq}) into the form
\begin{equation}
\Box_\eta {h}^{\mu\nu} =  -16\pi G_0 \tau^{\mu\nu} \,,
\label{heq0}
\end{equation}
where $\Box_\eta$ is the flat spacetime d'Alembertian with respect to $\eta_{\mu\nu}$, and where
\begin{align}
 \tau^{\mu\nu} &=  (-g) \left ( T^{\mu\nu} - K_{\rm ae}^\mu K_{\rm ae}^\nu K_{{\rm ae}\,\alpha} T_{\rm ae}^\alpha \right )  +  (-g) \left ( t_{LL}^{\mu\nu} + t_H^{\mu\nu} \right ) 
 \nonumber \\
 &
 \qquad +\frac{1}{8\pi G_0}(-g)S^{\mu\nu} \,,
\label{heq}
\end{align}
where   $t_H^{\mu\nu}$ is the Harmonic pseudotensor (see Eqs.\ (6.5) and (6.53) of \cite{PW2014} for explicit formulae for $(-g)t_{LL}^{\mu\nu}$ and $(-g)t_H^{\mu\nu}$). 

\section{Formal structure of the near-zone fields}
\label{sec:nearzonefields}

\subsection{Metric in terms of the fields}

 The next task will be to solve these equations iteratively in a post-Newtonian expansion in the near-zone, i.e. within one characteristic gravitational wavelength $\lambda$ of the center of mass of the system, in terms of a small parameter $\epsilon \sim v^2 \sim G_0 m/r$, where $v$, $r$ and $m$ are the characteristic velocities, separations and masses of the bodies in the system.  The strong-field internal gravity effects of each body will be encoded in expressions for the energy-momentum quantities  $T^{\mu\nu}$ and $T_{\rm ae}^{\nu} $.

We follow  \cite{2000PhRvD..62l4015P} (hereafter referred to as PWI) by defining a simplified notation for the field
${h}^{\mu\nu}$:
\begin{eqnarray}
\tilde{N} &\equiv& {h}^{00} \sim O(\epsilon) \,, \nonumber \\
\tilde{K}^j &\equiv& {h}^{0j} \sim O(\epsilon^{3/2}) \,, \nonumber \\
\tilde{B}^{jk} &\equiv& {h}^{jk} \sim O(\epsilon^2) \,, \nonumber \\
\tilde{B} &\equiv& {h}^{jj} \equiv \sum_j {h}^{jj} \sim O(\epsilon^2) \,.
\label{eq:fields}
\end{eqnarray}
We assume that the coordinate system is at rest with respect to the mean rest frame of the universe that is singled out by the asymptotic value of the {\AE}ther field $K_{\rm ae}^\mu$.  This implies that the asymptotic values of the spatial components vanish, and that therefore, within the near zone, they behave as

\begin{equation}
\tilde{K}_{\rm ae}^j \sim O(\epsilon^{3/2}) \,.
\label{eq:Kjform}
\end{equation}
Later, when we have the equations of motion and gravitational wave signals in hand, we will be able to transform them to a frame in which the system is at rest, using a suitably expanded Lorentz transformation combined with a gauge transformation (often called a ``post-Galilean'' transformation).

From the constraint on the norm of $K_{\rm ae}^\mu$ it follows that $K_{\rm ae}^0$ can be expressed in terms of the variables of Eqs.\ (\ref{eq:fields}) and (\ref{eq:Kjform}):
\begin{align}
K_{\rm ae}^0 &= 1+ \frac{\epsilon}{4}  \tilde{N} +  \frac{\epsilon^2}{4} \left (\tilde{B} - \frac{3}{8} \tilde{N}^2 \right )
+ \frac{\epsilon^3}{16}  \biggr [ \tilde{N} \tilde{B} + \frac{7}{8} \tilde{N}^3 
\nonumber \\
&\quad 
+ 8 \tilde{K}_{\rm ae}^j \tilde{K}_{\rm ae}^j- 16\tilde{K}^j \tilde{K}_{\rm ae}^j +4 \tilde{K}^j \tilde{K}^j \biggr ] +O(\epsilon^4) \,.
\end{align}
The harmonic gauge condition becomes $\tilde{N}_{,0}+ \tilde{K}^j_{,j} =0$ and $\tilde{K}^j_{,0}+\tilde{B}^{jk}_{,k} =0$.
Hereafter we do not distinguish between covariant and contravariant components of spatial indices, which are assumed to be raised or lowered using the Minkowski metric, whose spatial components are $\delta_{ij}$. 

In the equations of motion to 2.5PN order, we need to determine
the components of the physical metric and $\tilde{K}_{\rm ae}^j$ to the following orders:
$g_{00}$ to $O(\epsilon^{7/2})$,
$g_{0j}$ to $O(\epsilon^{3})$,
$g_{jk}$ to $O(\epsilon^{5/2})$, and 
$\tilde{K}_{\rm ae}^j$to $O(\epsilon^{3})$.
From the definitions (\ref{Hdef}) and  (\ref{hdef}), one can invert to find
$g_{\mu\nu}$ in terms of ${h}^{\mu\nu}$ to the appropriate order in $\epsilon$, as in PWI, Eq.\ (4.2).   
Expanding to the
required order, we find,
\begin{align}
g_{00} &= -1 +   \frac{\epsilon}{2} \tilde{N} 
+  \frac{\epsilon^2}{8}  \left ( 4\tilde{B} - 3 \tilde{N}^2 \right )
\nonumber \\
& \quad
 +  \frac{\epsilon^3}{16} \left ({5} \tilde{N}^3 - {4} \tilde{N}\tilde{B} +8 \tilde{K}^j \tilde{K}^j  \right )
 +O(\epsilon^4) \,, 
\nonumber\\
g_{0j} &= -  \epsilon^{3/2} \tilde{K}^j  +   \frac{\epsilon^{5/2}}{2} \tilde{N}  \tilde{K}^j  +O(\epsilon^{7/2}) \,, 
\nonumber \\
g_{jk} &= \delta^{jk} \left \{ 1+  \frac{\epsilon}{2} \tilde{N} -  \frac{\epsilon^2}{8}  \left ( \tilde{N}^2 + 4 \tilde{B}  \right ) \right \} 
\nonumber \\
&
\quad +  \epsilon^2 \tilde{B}^{jk} 
 +O(\epsilon^3) \,, 
\nonumber \\
(-g) &= 1+  \epsilon\tilde{N}  - \epsilon^2 \tilde{B} + O(\epsilon^3) \,.
\label{eq:metric}
\end{align}

\subsection{Change of field variables}

We can now use these definitions to express the field equations and the \AE ther equation to the required PN order in terms of $\tilde{N}$, $\tilde{K}^j$, $\tilde{B}^{jk}$, $\tilde{B}$ and $\tilde{K}_{\rm ae}^j$.  For example, the components of the  combination $t_{LL}^{\mu\nu} + t_H^{\mu\nu}$ have the same form as the components of $\Lambda^{\mu\nu}$ found in Eq.\ (4.4) of PWI.  However, despite the fact that the \AE ther energy-momentum tensor $S^{\mu\nu}$ is formally quadratic and higher-order in the fields, the fact that $K_{\rm ae}^0 = 1$ at lowest order implies that the fields $\tilde{N}$, $\tilde{K}^j$, $\tilde{K}_{\rm ae}^j$, $\tilde{B}^{jk}$ and $\tilde{B}$  can contribute {\em linearly} to the effective source of the relaxed field equation.  Even worse, the function $\tilde{N}$ contributes to $S^{00}$ at {\em Newtonian} order.  At purely linear order in the fields, the vacuum versions of the relaxed field equations take the form for the metric fields,
\begin{align}
\Box \tilde{N} &= \frac{1}{2} c_{14} \left [ \nabla^2 (\tilde{N}+\epsilon \tilde{B}) - 4 \epsilon ({\tilde{K}_{{\rm ae},0j}^j} - \tilde{K}^j_{,0j}  ) \right ]\,,
\nonumber \\
\Box \tilde{K}^j &= \frac{1}{2} c_2 \left [ 3 \tilde{N}_{,0j} + 4{{\tilde{K}}_{{\rm ae}\,,jk}^k}  -\epsilon \tilde{B}_{,0j}  \right] \,,
\nonumber \\
\Box \tilde{B}^{jk} &= -\frac{1}{2} c_2  \delta^{jk} \left [ 3 \tilde{N}_{,00}  + 4{\tilde{K}}^k_{{\rm ae},k0} - \epsilon \tilde{B}_{,00} \right ] \,,
\label{eq:linear1}
\end{align}
and for the \AE ther field,
\begin{align}
c_1 &\left [ \nabla^2 \left (\tilde{K}_{\rm ae}^j -\tilde{K}^j \right ) - \left (\tilde{K}_{\rm ae}^k -\tilde{K}^k \right )_{,kj} \right ]
\nonumber \\
&
  = -\frac{1}{4} c_{14} \left [ \tilde{N}_{,0j} +\epsilon \tilde{B}_{,0j} - 4 \epsilon\left (\tilde{K}_{\rm ae}^j -\tilde{K}^j \right )_{,00} \right ] 
\nonumber \\
& \quad
-\frac{1}{4} c_2 \left [ 3 \tilde{N}_{,0j}  + 4\tilde{K}^k_{{\rm ae},kj}  -\epsilon \tilde{B}_{,0j}\right ] \,.
\label{eq:linear2}
\end{align}
In Appendix \ref{app:waves}, we will study the wavelike solutions of these equations as an alternative method to verify the speeds and polarizations of waves derived in the literature \cite{2006PhRvD..73f4015F}.

These coupled, linear-order terms complicate the iteration procedure that is part of the post-Minkowskian method in general relativity or scalar-tensor theories, which rely upon the contributions to the right-hand-side of the relaxed equations being quadratic and higher in the small field quantities.  However, it turns out that a suitable change of variables eliminates these linear terms to the desired 2.5PN order; details are given in Appendix \ref{app:transform}.  This transformation is given by
\begin{align}
\tilde{N} &= {N} + \frac{\epsilon c_{14}}{2-c_{14}} B 
+ \frac{\epsilon}{v_L^2} \dot{R} 
+  \frac{\epsilon^2 c_{14}}{2(2-c_{14})v_L^2} \ddot{{X}}_B 
 \nonumber \\
 & \quad 
 + \frac{\epsilon^2}{12v_L^4} \left (  \stackrel{(3)}{{Y}_{R}} - c_1 v_L^2 W_T \stackrel{(3)}{{Y}}_{Kae} \right ) +O(\epsilon^3 N) \,,
 \nonumber \\
 \tilde{K}^j &= K^j  - {R}_{,j} - \frac{\epsilon c_{14}}{2(2-c_{14})} \dot{X}_{B,j} 
 \nonumber \\
 & \quad 
 - \frac{\epsilon}{12 v_L^2}
 \left (  \ddot{Y}_{{R},j} - c_1 v_L^2 W_T  \ddot{Y}_{Kae,j} \right ) +O(\epsilon^2 K^j)\,,
 \nonumber \\
 \tilde{B}^{jk} &= {B}^{jk} +  \delta^{jk} \biggl [ \dot{R} + 
 \frac{\epsilon c_{14}}{2(2-c_{14})} \ddot{{X}}_B  
 \nonumber \\
 & \quad 
 +\frac{\epsilon}{12v_L^2} \left (  \stackrel{(3)}{{Y}_{R}} - c_1 v_L^2 W_T \stackrel{(3)}{{Y}}_{Kae} \right ) \biggr ] +O(\epsilon^2 B^{jk})\,,
 \nonumber \\
 \tilde{K}_{\rm ae}^j &= K_{\rm ae}^j + K^j  +\frac{1}{2c_{14}} \left (W_L {R}_{,j}
 +  c_1 W_T X_{Kae,j} \right ) 
 \nonumber \\
 & \quad 
 + \frac{\epsilon}{4} \frac{W_L}{2-c_{14}} \dot{X}_{B,j}
  + \frac{\epsilon}{24 c_{14} v_L^2}
 \biggl ( W_L  \ddot{Y}_{{R},j} 
  \nonumber \\
  & \quad
  + c_1 v_L^2 (1-W_L)W_T  \ddot{Y}_{Kae,j} \biggr ) +O(\epsilon^2 K_{\rm ae}^j)
 \,,
 \label{eq:transform}
\end{align}
where
\begin{align}
v_T^2 & \equiv \frac{c_1}{c_{14}}  \,,
\nonumber \\
v_L^2 & \equiv \frac{c_2 (2-c_{14})}{c_{14} (2+3c_2)} \,,
 \end{align}
 are the propagation speeds of the transverse and longitudinal waves of the \AE ther field (Appendix \ref{app:waves}), and where
 \begin{equation}
W_T \equiv 1-\frac{1}{v_T^2} \,, \quad W_L \equiv \left (1-\frac{c_{14}}{2} \right ) \left (1-\frac{1}{v_L^2} \right ) \,.
\end{equation}

Here and for future use, we define an array of ``superpotentials'' $X$, superduperpotentials $Y$ and ``megasuperpotentials'' $Z$, defined by
\begin{align}
\nabla^2 {X}_N& \equiv 2 {N} \,, \quad \nabla^2 {Y}_N \equiv 12 X_N \,,\quad \nabla^2 Z_N \equiv 30 Y_N \,, 
\nonumber \\
\nabla^2 {X}_{K{\rm ae}} &\equiv 2 {K}^k_{{\rm ae},k} \,,\nabla^2 {Y}_{K{\rm ae}} \equiv 12 X_{K{\rm ae}} \,,
\nonumber \\
& \qquad  \nabla^2 Z_{K{\rm ae}} \equiv 30 Y_{K{\rm ae}} \,, 
 \nonumber\\
\nabla^2 {X}_B &\equiv 2 {B} \,,\nabla^2 {Y}_B \equiv 12 X_B \,,\nabla^2 Z_B \equiv 30 Y_B \,,
\label{eq:super}
\end{align}
along with the superpotential combination,
\begin{equation}
{R} \equiv \frac{1}{4} c_{14} \dot{X}_N - c_1 X_{K{\rm ae}} \,,
\end{equation}
and its own superduperpotential defined by $\nabla^2 Y_{R} = 12 {R}$.

The harmonic gauge conditions $\tilde{N}_{,0}+ \tilde{K}^j_{,j} =0$ and $\tilde{K}^j_{,0}+\tilde{B}^{jk}_{,k} =0$ imply gauge conditions in the new variables given by 

\begin{align}
K^j_{,j} + \left (1- \frac{c_{14}}{2} \right )\dot{N} &= - 2c_1 K^j_{{\rm ae},j}  - \epsilon c_1 W_T \ddot{X}_{\rm ae}  +O(\epsilon^2 \dot{N})\,,
\nonumber \\
\dot{K}^j + {B^{jk}}_{,k} &= O(\epsilon \dot{K}^j) \,.
\label{eq:newgauge}
\end{align}

\subsection{Final relaxed \eae equations}

In terms of the new variables, the relaxed \eae  equations take the form
\begin{subequations}
\begin{align}
  \left (1-\frac{1}{2}c_{14}\right) \Box N &= -16\pi G_0 \tau^{00} + O(\rho \epsilon^3) \,,
\label{eq:fieldeqsnewN} \\
\Box K^j &= -16\pi G_0 \tau^{0j}  + O(\rho \epsilon^{5/2}) \,,
\label{eq:fieldeqsnewV} \\
\Box B^{jk} &= -16\pi G_0 \tau^{jk}  + O(\rho \epsilon^{2}) \,,
\label{eq:fieldeqsnewBjk} \\
\Box B &= -16\pi G_0  \tau^{kk}+ O(\rho \epsilon^{3}) \,,
\label{eq:fieldeqsnewB} \\
c_1 \Box^* K_{\rm ae}^j &= 8 \pi G_0 \tau_{\rm  ae}^j  + O(\rho \epsilon^{5/2}) \,,
\label{eq:fieldeqsnewK}
\end{align}
\label{eq:fieldeqsnew}
\end{subequations}
where $\Box^* \equiv \nabla^2 - v_T^{-2} \partial_0^2$, and where
\begin{align}
\tau^{\mu\nu} &\equiv (-g) T_T^{\mu\nu} + (16\pi G_0)^{-1} \Lambda_T^{\mu\nu} \,,
\nonumber \\
\tau_{\rm ae}^j & \equiv T_{{\rm  ae} \, T}^j  + (8 \pi G_0)^{-1} \Lambda^j_{\rm ae} \,.
\label{eq:tau}
\end{align}

In obtaining these equations, we made use of the spatial components of the  {\AE}ther field equation (\ref{eq:aether2}) to make further simplifications of the $S^{\mu\nu}$.  Pulling all the matter contributions together gives the total matter source tensor,
\begin{align}
T_T^{00} &= T^{00} - (K_{\rm ae}^0)^2 K_{{\rm ae}\,\mu} T_{\rm ae}^\mu 
\nonumber \\
& \quad + 2 (\tilde{K}_{\rm ae}^j - \tilde{K}^j)(T_{\rm ae}^j + \tilde{K}_{\rm ae}^j K_{{\rm ae}\,\mu} T_{\rm ae}^\mu )
\,,
\nonumber \\
T_T^{0j} & = T^{0j}  + K_{\rm ae}^0 T_{\rm ae}^j \,,
\nonumber \\
T_T^{jk} & = T^{jk} + 2\tilde{K}_{\rm ae}^{(j} T_{\rm ae}^{k)} + \tilde{K}_{\rm ae}^j \tilde{K}_{\rm ae}^k  K_{{\rm ae}\,\mu} T_{\rm ae}^\mu \,,
\nonumber \\
 T_{{\rm  ae} \, T}^j &= T_{\rm  ae}^j + \tilde{K}_{\rm ae}^j K_{{\rm ae} \, \mu} T_{\rm ae}^\mu \,,
\label{eq:TT}
\end{align}
where $K_{{\rm ae} \, \mu}$ is a four vector with components $(K_{{\rm ae} \,0},\tilde{K}_{{\rm ae} \, j})$.
Many of these contributions to  $T_T^{\mu\nu}$ are of higher PN order than we need, but we will address this when we introduce our ``compact point mass model'' for the matter sources.  We will also transform all potentials in Eq.\ (\ref{eq:TT}) such as $\tilde{K}^j$ and $\tilde{K}_{\rm ae}^j$ to our new potentials using Eqs.\ (\ref{eq:transform}). 
The field contributions to $S^{\mu\nu}$ have been combined with the Landau-Lifshitz and Harmonic pseudotensors to produce the total $\Lambda_T^{\mu\nu}$ and $\Lambda_{\rm ae}^j$, given by
\begin{widetext}    
\begin{subequations} 
\begin{align}
\Lambda_T^{00} &= - \frac{7}{16} (2 -  c_{14}) {N_{,j}}^2
\nonumber \\
& \quad
+\frac{7}{16} (2 -  c_{14}) N {N_{,j}}^2 + \frac{1}{4} (1 - 4  c_{14}) N_{,j} B_{,j} - \frac{1}{2} (2 -  c_{14}) B_{jk} N_{,jk} 
 - \frac{1}{4} \left [ (2-c_{14}) ( 2 + 3c_{14}) - c_{14} W_L \right] N \ddot{N} 
\nonumber \\
& \quad
 + \frac{5}{16} \left \{ (2-c_{14})(1-2 c_{14}) +  c_{14} W_L \right \} \dot{N}^2 
  + (1 +  c_{14}) N_{,j} \dot{K}^j
  -\frac{1}{2}( 4  - 5 c_{14}) \dot{N}_{,j} K^j 
   + \frac{1}{2} K^j_{,k} (K^j_{,k} + 3 K^k_{,j}) 
  \nonumber \\
& \quad 
   + \frac{3}{2}  c_{14} N_{,j} \dot{K}_{\rm ae}^j 
+ \frac{3}{{2}}  c_{14}   \dot{N}_{,j} K_{\rm ae}^j
 -  \frac{1}{2} c_{14} N \nabla^2 B
  + 2  c_{14} K^k_{{\rm ae},j} (K^j_{,k} + K^j_{{\rm ae},k})
  + 6 c_1 K_{\rm ae}^{[j,k]} K_{\rm ae}^{[j,k]} 
  +  \frac{1}{2} c_{14} \nabla^2 (K^j K^j)
 \nonumber \\
& \quad   
  +  c_{14} \left (K_{\rm ae}^j \nabla^2 K^j  - K^j \nabla^2 K_{\rm ae}^j + \nabla^2 (K^j  K_{\rm ae}^j) \right )
+\frac{1}{4} c_1 \left ( 6-12 c_{14}- 4W_L -3W_T \right ) N \dot{K}^j_{{\rm ae},j}
  \nonumber \\
& \quad 
+\frac{1}{2} \frac{ c_1}{c_{14}} \left ( 1-2c_{14} - W_L \right ) (3c_{14} \dot{N} - 2c_1  {K}^j_{{\rm ae},j}) {K}^j_{{\rm ae},j}
-2c_1 (1+W_T) {K}^j_{{\rm ae}} {K}^k_{{\rm ae},kj}
-2c_1 (1+W_T) K^j {K}^k_{{\rm ae},kj} 
\nonumber \\
& \quad 
+ \frac{1}{4} \left (4(2 - c_{14}) + 3W_L \right ) \dot{N}_{,j} {R}_{,j}
+\frac{3}{4}c_1 W_T  \dot{N}_{,j} X_{Kae,j}
- \frac{c_1}{c_{14}}  W_L (1+W_T) {K}^k_{{\rm ae},kj} {R}_{,j}
 \nonumber \\
& \quad
-\frac{c_1^2}{c_{14}} W_T (1+W_T) {K}^k_{{\rm ae},kj} X_{Kae,j}
 -\frac{5}{4} W_L  \dot{R} \nabla^2 N
 -\frac{3}{8} c_1 W_T \dot{X}_{Kae} \nabla^2 N 
 +2{R}_{,j}  \nabla^2 K^j 
\nonumber \\
& \quad
- (W_L {R}_{,j} + c_1 W_T X_{Kae,j}) \nabla^2 K^j_{{\rm ae}} 
-\frac{1}{4} \left ( 4- 2c_{14} - 5W_L\right )  \nabla^2 (N \dot{R} )
+\frac{3}{8} c_1 W_T \nabla^2 (N \dot{X}_{Kae})
\nonumber \\
& \quad
- (2 -c_{14} ) \nabla^2 \left ( K^j  {R}_{,j}  \right ) 
+ \nabla^2 \left [\left (K^j + K^j_{{\rm ae}} \right ) \left (W_L {R}_{,j} + c_1 W_T X_{Kae,j}\right ) \right ]
\nonumber \\
& \quad
+ \frac{1}{2}  (2-c_{14}) \nabla^2 \left ( {R}_{,j}{R}_{,j}\right) + \frac{1}{4c_{14}} \nabla^2 \left | W_L{R}_{,j} +c_1 W_T X_{Kae,j} \right |^2
  + O(\rho \epsilon^3 ) \,,
\\
\Lambda_T^{0j} &= 
 N_{,k} \left [ (1 -  c_{2}) K^k_{,j} - K^j_{,k} \right ]
+ \frac{3}{8} (2 + 3  c_{2})  \dot{N} N_{,j} 
+ \frac{1}{4} c_2 (6-3c_{14} - W_L) N \dot{N}_{,j}   - \frac{1}{2} c_{14} \nabla^2 N (K^j + K_{\rm ae}^j) 
\nonumber \\
& \quad
+ \frac{1}{2} N_{,k} \left [  ( c_{14} - 2  c_{2}) K^k_{{\rm ae},j} -c_{14} K^j_{{\rm ae},k}\right ] -  c_{2} N_{,jk} (K^k + K_{\rm ae}^k) 
- \frac{1}{2} \frac{c_1c_2}{c_{14}} (3-6c_{14} - 2W_L-W_T)N K^k_{{\rm ae},kj} 
\nonumber \\
& \quad
- \frac{c_2}{2c_{14}} N_{,jk} (W_L {R}_{,k} + c_1 W_T X_{Kae,k} )
+ \frac{1}{4c_{14}} \left (c_2 -c_{14} \right ) \nabla^2 N\, (W_L {R}_{,j} + c_1 W_T X_{Kae,j} ) 
\nonumber \\
& \quad
- \frac{c_2}{4c_{14}} \nabla^2 \left \{ N (W_L {R}_{,j} + c_1 W_T X_{Kae,j} ) \right \} 
+ O(\rho \epsilon^{5/2} )\,,
\\
\Lambda_T^{jk} &= \frac{1}{8} (2 -  c_{14}) \left [ N_{,j} N_{,k} - \frac{1}{2} \delta^{jk} {N_{,m}}^2 \right ] 
+ O(\rho \epsilon^2 ) \,,
\\
\Lambda_T^{jj} &=
-\frac{1}{16} (2 -  c_{14})  {N_{,j}}^2
\nonumber \\
& \quad
+\frac{1}{8} (2 -  c_{14}) N {N_{,j}}^2
 - \frac{1}{4} N_{,j} B_{,j} 
+ \frac{3}{16c_2}  \left  (c_{14}^2-18 c_2^2- 6 c_2 \right )  \dot{N}^2 
- (1- 3  c_{2})  N_{,j} \dot{K}^j 
-  K^{[j,k]} K^{[j,k]} 
\nonumber \\
& \quad
+\frac{3}{4} \left (c_{2} c_{14} - c_{14} -5c_{2} \right )  N \ddot{N} 
 +\frac{1}{2} (6 c_2 - c_{14}) \dot{N}_{,j} (K^j + K_{\rm ae}^j)
  +\frac{1}{2} (6 c_2 -c_{14}) {N}_{,j} \dot{K}_{\rm ae}^j
\nonumber \\
& \quad 
+ 2 c_1 K_{\rm ae}^{[j,k]} K_{\rm ae}^{[j,k]} 
+\frac{1}{4} \left ( 13c_1 -12c_1 c_2  -c_{14} + 6c_2 \right )N \dot{K}^j_{{\rm ae},j} 
- \frac{3c_{14} c_1}{2  c_{2}} \dot{N} K^j_{{\rm ae},j}
+ 2c_1 (K^j + K_{\rm ae}^j) K^k_{{\rm ae},kj} 
\nonumber \\
& \quad
+  \frac{3c_1^2}{c_{2}}  (K^k_{{\rm ae},k})^2  
+ \frac{1}{4 c_{14}} \left ( 6c_2- c_{14} \right ) \dot{N}_{,j} \left ( W_L {R}_{,j} + c_1 W_T X_{Kae,j} \right )
+ \frac{c_1}{c_{14}} K^k_{{\rm ae},kj} \left ( W_L {R}_{,j} + c_1 W_T X_{Kae,j} \right )
\nonumber \\
& \quad
+ \frac{3 c_2}{4 c_{14}} W_L \left ( \nabla^2 (N \dot{R} )- \dot{R} \nabla^2 N \right )
+\frac{c_1}{8c_{14}}  \left ( 6c_2 -c_{14} \right ) W_T \left ( \nabla^2 (N \dot{X}_{Kae} ) - \dot{X}_{Kae} \nabla^2 N \right )
+ O(\rho \epsilon^3 ) \,,
\\
\Lambda_{\rm ae}^{j} &=
\frac{1}{16} (3 c_2+c_{14}-3 c_1) \left (4- 2c_{14} -W_L \right ) \dot{N} N_{,j} 
- \frac{c_1}{8c_{14}} \left ( 12c_1 c_{14} +(3c_1 - c_{14} + 2c_2 )(W_L - W_T) \right ) N K^k_{{\rm ae},kj} 
\nonumber \\
& \quad
+ \frac{1}{32} \left \{ 6(c_{14} + 3c_2 )-12 c_1 (2 - c_{14} )+(3c_1-c_{14}+2c_2)W_L \right \} N \dot{N}_{,j} 
- \frac{c_1}{4  c_{2}} ( c_{14} - 3 c_1 ) N_{,j} K^k_{{\rm ae},k} 
\nonumber \\
& \quad
+ \frac{1}{4} (3 c_1 -  c_{14} - 2  c_{2}) N_{,jk} (K^k + K_{\rm ae}^k) 
+ \frac{1}{4} ( c_{14} - c_1 - 2  c_{2}) N_{,k} K^k_{{\rm ae},j} - \frac{1}{2} c_1 N_{,k} K^j_{{\rm ae},k} 
\nonumber \\
& \quad
+ \frac{1}{4} (3 c_1 - 2  c_{2}) N_{,k} K^k_{,j} 
- \frac{1}{4} (6c_1 - c_{14} ) N_{,k} K^j_{,k} - \frac{3}{4} c_1 \nabla^2 N (K^j + K_{\rm ae}^j)  
+ \frac{1}{4} c_1N \nabla^2  (2 K_{\rm ae}^j - 3 K^j)
\nonumber \\
& \quad
+\frac{1}{8c_{14}} (3c_1-c_{14}-2c_2) N_{,jk} (W_L {R}_{,k} + c_1 W_T X_{Kae,k} )
-\frac{1}{16c_{14}} (3c_1+c_{14}-2c_2) \nabla^2 N(W_L {R}_{,j} + c_1 W_T X_{Kae,j} )
\nonumber \\
& \quad
-\frac{1}{16c_{14}} (3c_1-c_{14}+2c_2) \nabla^2 [N(W_L {R}_{,j} + c_1 W_T X_{Kae,j} )]
+ O(\rho \epsilon^{5/2} ) \,.
\end{align}
\label{eq:LambdaT}
\end{subequations}

As is customary, we will define the quantities
\begin{align}
\sigma &\equiv T_T^{00} + T_T^{jj} \,,
\nonumber \\
\sigma^j &\equiv T_T^{0j}  \,,
\nonumber \\
\sigma^{jk} &\equiv T_T^{ik}  \,,
\nonumber \\
\sigma_{\rm ae}^j &\equiv T_{{\rm ae}\, T}^j \,,
\end{align}
and will express various potentials formally in terms of these densities.  Later we will make a PN expansion of them, including compact body sensitivities, and iterate the potentials to include these effects.

The gauge conditions (\ref{eq:newgauge}) lead to useful conservation equations.  Adding the time derivative of Eq.\ (\ref{eq:fieldeqsnewN}) to the divergence of Eq.\ (\ref{eq:fieldeqsnewV}), and making use of the d'Alembertian of the first of Eqs.\ (\ref{eq:newgauge}) and the divergence of Eq.\ (\ref{eq:fieldeqsnewK}), we obtain the equations, valid through 2PN order
\begin{align}
{\tau^{0\nu}}_{,\nu} - \tau^j_{{\rm ae},j} &=0\,,
\nonumber \\
{\tau^{j\nu}}_{,\nu} &= 0 \,. 
\label{eq:conservation}
\end{align} 

\subsection{Near-zone field to 2.5PN order}

We now solve Eqs.\ (\ref{eq:fieldeqsnew}) for field points within the near-zone.  The formal solutions of Eqs.\ (\ref{eq:fieldeqsnew}) consist of integrals of the source divided by $|\bm{x}-\bm{x}'|$ over the past harmonic ``null'' cone of the field point.  These integrals divide into two distinct integrals, an inner integral out to a boundary where the null cone $\cal C$ intersects the near-zone world tube of radius ${\cal R} \sim \lambda$, and an outer integral over the remainder of the null cone.  The retarded time of the inner integrals over the region $\cal N$ can be expanded in powers of $|\bm{x}-\bm{x}'|$, leading to bounded integrals over a constant time hypersurface $\cal M$, evaluated at the time $t$ of the field point.  The outer integrals over the rest of the null cone ${\cal C} - {\cal N}$ are carried out using a special change of integration variables.   For a detailed pedagogical description of  this method, see Sec. 6.3 of \cite{PW2014}.  Both the inner and outer integrals may individually depend on the radius ${\cal R}$, but their sum cannot; in practice this means that one can evaluate each integral, keeping only terms that do not depend explicitly on $\cal R$.

The expansions of the inner integrals of the fields $N$, $K^j$, $B^{jk}$ and $K_{\rm ae}^j$ are then given by
\begin{subequations}
\label{bigexpansion}
\begin{eqnarray}
N_{\cal N} &=& \frac{2G_0}{2-c_{14}} \biggl [4 \epsilon \int_{\cal M} \frac{\tau^{00}(t,{\bf x}^\prime)}{|{\bf x}-{\bf
x}^\prime |} d^3x^\prime 
+2 \epsilon^2 \partial^2_t \int_{\cal M} \tau^{00}(t,{\bf x}^\prime) |{\bf x}-{\bf x}^\prime | d^3x^\prime
-\frac{2}{3} \epsilon^{5/2} \stackrel{(3)\qquad}{{\cal I}^{kk}(t)} 
-\frac{4}{3}  \epsilon^{5/2} x^k \ddot{\cal I}_{\rm ae}^k(t)
\nonumber \\
&&
+ \frac{1}{6} \epsilon^3 \partial^4_t \int_{\cal M}
\tau^{00}(t,{\bf x}^\prime) |{\bf x}-{\bf x}^\prime |^3 d^3x^\prime 
- \frac{1}{30} \epsilon^{7/2} \biggl \{ (4x^{kl}+2r^2\delta^{kl})
\stackrel{(5)\quad}{{\cal I}^{kl}(t)}
- 4 x^k \stackrel{(5)\qquad}{{\cal I}^{kll}(t)}
+ \stackrel{(5)\qquad}{{\cal I}^{kkll}(t)} \biggr \} 
\nonumber \\
&&
-\frac{2}{15}  \epsilon^{7/2} r^2 x^k  \stackrel{(4)\qquad}{{\cal I}_{\rm ae}^{k}(t)}
+ N_{\partial {\cal M}} \biggr ]+ O(\epsilon^4) \,,
\label{bigexpansiona} \\
K^j_{\cal N} &=& G_0 \biggl [4 \epsilon^{3/2} \int_{\cal M} \frac{\tau^{0j}(t,{\bf x}^\prime)}{|{\bf x}-{\bf
x}^\prime |} d^3x^\prime 
+2 \epsilon^{5/2} \partial^2_t \int_{\cal M} \tau^{0j}(t,{\bf x}^\prime) |{\bf x}-{\bf x}^\prime | d^3x^\prime
 +\frac{2}{9} \epsilon^3 \biggl \{ 3 x^k \stackrel{(4)\quad}{{\cal I}^{jk}(t)}
- \stackrel{(4)\qquad}{{\cal I}^{jkk}(t)}
+2 \epsilon^{mjk}  \stackrel{(3)\qquad}{{\cal J}^{mk}(t)}
\biggr \} 
\nonumber \\
&&
+\frac{2}{9} \epsilon^3 \biggl \{ 6 x^k \stackrel{(3)\qquad}{{\cal I}_{\rm ae}^{(jk)}(t)}
- \stackrel{(3)\qquad}{{\cal I}_{\rm ae}^{jkk}(t)} -2 \stackrel{(3)\qquad}{{\cal I}_{\rm ae}^{kjk}(t)} \biggr \}
+ K^j_{\partial {\cal M}} \biggr ]+ O(\epsilon^{7/2}) \,,
\label{bigexpansionb} \\
B^{jk}_{\cal N} &=&G_0 \biggl [ 4 \epsilon^2 \int_{\cal M} \frac{\tau^{jk}(t,{\bf x}^\prime)}{|{\bf x}-{\bf x}^\prime |} d^3x^\prime 
- 2 \epsilon^{5/2} \biggl \{ \stackrel{(3)\quad}{{\cal I}^{jk}(t)} +2\ddot{\cal I}_{\rm ae}^{(jk)}(t) \biggr \}
+2 \epsilon^3 \partial^2_t \int_{\cal M}
\tau^{jk}(t,{\bf x}^\prime) |{\bf x}-{\bf x}^\prime | d^3x^\prime
\nonumber \\
&& - \frac{1}{9} \epsilon^{7/2} \biggl \{ 
3 r^2 \stackrel{(5)\quad}{{\cal I}^{jk}(t)}
-2x^l \stackrel{(5)\qquad}{{\cal I}^{jkl}(t)}
- 8 x^l \epsilon^{ml(j} \stackrel{(4)\qquad}{{\cal J}^{m|k)}(t)}
+ 6 \stackrel{(3)\qquad}{M^{jkll}(t)} \biggr \}
\nonumber \\
&& - \frac{2}{3} \epsilon^{7/2} \biggl \{ 
 r^2 \stackrel{(4)\quad}{{\cal I}_{\rm ae}^{(jk)}(t)}
+x^l \stackrel{(4)\qquad}{{\cal I}_{\rm ae}^{(jkl)}(t)}
\biggr \}
 + B^{jk}_{\partial {\cal M}} \biggr ]+ O(\epsilon^4) \,,
\label{bigexpansionc}\\
K^j_{{\rm ae}\,{\cal N}} &=& - G_0 c_1^{-1} \biggr [ 2 \epsilon^{3/2} \int_{\cal M} \frac{\tau_{\rm ae}^j(t,{\bf x}^\prime)}{|{\bf x}-{\bf x}^\prime |} d^3x^\prime 
- \frac{2}{v_T} \epsilon^{2} \dot{\cal I}_{\rm ae}^j  
+ \frac{1}{v_T^2} \epsilon^{5/2} \partial^2_t \int_{\cal M} \tau_{\rm ae}^j (t,{\bf x}^\prime) |{\bf x}-{\bf x}^\prime | d^3x^\prime
\nonumber 
\\
&&
-\frac{1}{3v_T^3} \epsilon^{3} \biggl \{ r^2 \stackrel{(3)\quad}{{\cal I}_{\rm ae}^j(t)}
-2x^k \stackrel{(3)\quad}{{\cal I}^{jk}_{\rm ae}(t)}
+ \stackrel{(3)\quad}{{\cal I}^{jkk}_{\rm ae}(t)} \biggr \} \biggr ]
 + O(\epsilon^{7/2}) \,,
 \label{bigexpansiond}
\end{eqnarray}
\end{subequations}
\end{widetext}
where we define the moments of the system by

\begin{eqnarray}
{\cal I}^Q &\equiv&  \int_{\cal M} \tau^{00} x^Q d^3x \,,
\nonumber
\\
{\cal J}^{iQ} &\equiv&  \epsilon^{iab}\int_{\cal M} \tau^{0b} x^{aQ} d^3x \,,
\nonumber
\\
M^{ijQ} &\equiv& \int_{\cal M} \tau^{ij} x^Q d^3x \,,
\nonumber
\\
{\cal I}_{\rm ae}^{jQ} &\equiv& \int_{\cal M} \tau^j_{\rm ae} x^Q d^3x \,.
\label{eq:genmoment}
\end{eqnarray}

The index $Q$ is a
multi-index, such that $x^Q$ denotes $x^{i_1} \dots x^{i_q}$.
The integrals are taken over a constant time hypersurface $\cal M$ at time $t$ out to the radius $\cal R$.
The structure of the expansions for ${N}_{\cal N}$, 
${K}^j_{\cal N}$ and ${B}^{jk}_{\cal N}$ differs from the structure in PWI only in the odd-half order $\epsilon$ terms in
${N}_{\cal N}$ and ${B}^{jk}_{\cal N}$ and in the integer-order $\epsilon$ terms in 
${K}^j_{\cal N}$.  This is because the source $\tau^{\mu\nu}$ satisfies the conservation law ${\tau^{\mu\nu}}_{,\nu}= \delta^\mu_0 \tau^j_{\rm ae, j}$ [Eq.\ (\ref{eq:conservation})], and we have used this to convert a number of terms into surface integrals at the boundary $\partial {\cal M}$ of the near zone.  In some places this leaves a residue of terms, such as the term proportional to $\epsilon^{5/2} x^k  \ddot{\cal I}_{\rm ae}^k(t)$ in ${N}_{\cal N}$.  Interestingly, while the terms in ${N}_{\cal N}$ and ${B}_{\cal N}$ that are proportional to $\epsilon^{5/2}$ are nominally of $1.5$PN order, we will see below that they are purely a gauge artifact, and do not contribute to the final metric at $1.5$PN order.  

The boundary terms $N_{\partial {\cal M}}$, $K^i_{\partial {\cal M}}$ and $B^{ij}_{\partial {\cal M}}$ can be found in Appendix C of PWI, with the replacement $\tau^{0j} \to  \tau^{0j}  - \tau^j_{\rm ae}$, but they will play no role in our analysis.  As in PWI, we will discard all terms that depend on the radius $\cal R$ of the near-zone; these necessarily cancel against terms that arise from integrating over the remainder of the past null cone.   As in general relativity, those ``outer'' integrals can be shown to make no contribution to the near zone metric to the 2.5PN order at which we are working (see Sec.\ 4.C of PWI) .

\subsection{Definitions of PN and 2PN potentials}
\label{sec:defpotentials}

In the near zone, the potentials are Poisson-like potentials and their generalizations.  Most were defined in PWI, but we will need to define additional potentials associated with the {\AE}ther field.  For a source
$f$, we define the Poisson potential to be
\begin{equation}
P(f) \equiv \frac{1}{4\pi} \int_{\cal M} \frac{f(t,{\bf x}^\prime)}
{|{\bf x}-{\bf x}^\prime | } d^3x^\prime \,, \quad \nabla^2
P(f) = -f \,.
\label{eq:Pf}
\end{equation}
We also define potentials based on the ``densities'' $\sigma$,
$\sigma^j$ and $\sigma^{jk}$ and $\sigma_{\rm ae}^j$: 
\begin{eqnarray}
\Sigma (f) &\equiv& \int_{\cal M} \frac{\sigma(t,{\bf x}^\prime)f(t,{\bf x}^\prime)}{|{\bf x}-{\bf x}^\prime | } d^3x^\prime = P(4\pi\sigma f) \,,
\nonumber\\
\Sigma^j (f) &\equiv& \int_{\cal M} \frac{\sigma^j(t,{\bf x}^\prime)f(t,{\bf
x}^\prime)}{|{\bf x}-{\bf x}^\prime | } d^3x^\prime = P(4\pi\sigma^j f) \,,
\nonumber\\
\Sigma^{jk} (f) &\equiv& \int_{\cal M} \frac{\sigma^{jk}(t,{\bf x}^\prime)f(t,{\bf x}^\prime)}{|{\bf x}-{\bf x}^\prime | } d^3x^\prime = P(4\pi\sigma^{jk} f) \,,
\nonumber\\
\Sigma_{\rm ae}^j (f) &\equiv& \int_{\cal M} \frac{\sigma_{\rm ae}^j (t,{\bf x}^\prime)f(t,{\bf
x}^\prime)}{|{\bf x}-{\bf x}^\prime | } d^3x^\prime = P(4\pi\sigma_{\rm ae}^j f) \,, 
\nonumber \\
\end{eqnarray}
along with the superpotentials
\label{definesuper}
\begin{eqnarray}
X(f)  &\equiv& \int_{\cal M} {\sigma(t,{\bf x}^\prime)f(t,{\bf
x}^\prime)}
{|{\bf x}-{\bf x}^\prime | } d^3x^\prime  \,,
\nonumber \\
Y(f) &\equiv& \int_{\cal M} {\sigma(t,{\bf x}^\prime)f(t,{\bf x}^\prime)}
{|{\bf x}-{\bf x}^\prime |^3 } d^3x^\prime  \,,
\end{eqnarray}
and their obvious counterparts $X^j$,  $X^{jk}$, $X_{\rm ae}^j$,  and so on.   Using Eq. (\ref{eq:Pf}), we can express the superpotential defined in Eq. (\ref{eq:super}) in the form
\begin{equation}
X_N = -2  P(N) \,, \, Y_N = -12 P(X_N) \,, \, Z_N = -30 P(Y_N) \,,
\end{equation}
and so on.

A number of potentials occur sufficiently frequently in the PN
expansion that it is useful to define them specifically.  There is the ``Newtonian'' potential,
\begin{eqnarray}
U \equiv \int_{\cal M} \frac{\sigma(t,{\bf x}^\prime)}{|{\bf x}-{\bf x}^\prime | } d^3x^\prime = P(4\pi\sigma) =
\Sigma(1) \,.
\end{eqnarray}
The potentials needed for the post-Newtonian limit are:
\begin{eqnarray}
V^j &\equiv& \Sigma^j(1) \,, \quad  V_{\rm ae}^j \equiv \Sigma_{\rm ae}^j (1) \,,  
\nonumber \\
\quad \Phi_1^{jk}  &\equiv& \Sigma^{jk}(1) \,, \quad
 \Phi_1 \equiv \Sigma^{jj}(1) \,, \quad \Phi_2 \equiv \Sigma(U)  \,,
 \nonumber \\
\quad X &\equiv& X(1) = -2P(U)\,, 
 \nonumber \\
\quad
X^j_{\rm ae} &\equiv& X^j_{\rm ae} (1) = -2P(V_{\rm ae}^j) \,.
\end{eqnarray}
Useful 2PN potentials include:
\begin{eqnarray}
&V_2^j \equiv \Sigma^j(U) \,, \qquad & V_{2{\rm ae}}^j \equiv \Sigma_{\rm ae}^j(U) \,,
\nonumber \\
&\Phi_2^j \equiv \Sigma(V^j) \,, \qquad & \Phi_{2{\rm ae}}^j \equiv \Sigma(V_{\rm ae}^j) \,,
\nonumber \\
& X_1 \equiv X^{jj}(1) = -2P(\Phi_1) \,, &X_2 \equiv  X(U) = - 2P(\Phi_2) \,,
\nonumber
\\
&X^j \equiv X^j(1) = -2P(V^j) \,, \quad & Y \equiv Y(1) \,,
 \nonumber \\
&P_2^{ij} \equiv P(U^{,i}U^{,j}) \,, \qquad & P_2 \equiv P_2^{ii}=\Phi_2
-\frac{1}{2}U^2 \,,
\nonumber \\
&G_1 \equiv P({\dot U}^2)  \,,
\qquad &G_2 \equiv P(U {\ddot U})  \,, 
\nonumber \\
&G_3 \equiv -P({\dot U}^{,k} V^k) \,, \qquad & G_{3{\rm ae}} \equiv
 -P({\dot U}^{,k} V_{\rm ae}^k) \,,
 \nonumber \\
&G_4 \equiv P(V^{i,j}V^{j,i}) \,, \qquad & G_{4{\rm ae}} \equiv P(V_{\rm ae}^{i,j}V^{j,i}) \,,
\nonumber \\
&G_{4{\rm ae}}^{\rm ae} \equiv P(V_{\rm ae}^{i,j}V_{\rm ae}^{j,i}) \,,
\nonumber \\
&G_5 \equiv -P({\dot V}^k U^{,k}) \,,  \qquad &G_{5{\rm ae}} \equiv -P({\dot V}_{\rm ae}^k U^{,k})
\nonumber \\
&G_6 \equiv P(U^{,ij} \Phi_1^{ij}) \,, 
\nonumber \\
&G_7^i \equiv P(U^{,k}V^{k,i}) \,,  \qquad &G_{7{\rm ae}}^i \equiv P(U^{,k}V_{\rm ae}^{k,i}) \,,
\nonumber \\
&G_8^i \equiv P(U^{,i}\dot U ) \,, \qquad & G_9^i \equiv P(U \dot{U}^{,i}) \,,
 \nonumber \\
& H \equiv P(U^{,ij} P_2^{ij}) \,. 
\label{potentiallist}
\end{eqnarray}

\section{Expansion of near-zone fields to 2.5PN order}
\label{sec:2.5expansion}

In evaluating the contributions at each order, we shall use the
following notation,
\begin{eqnarray}
N &=&  N_0 + \epsilon N_1+ \epsilon^{3/2} N_{1.5}+ \epsilon^2 N_2+
\epsilon^{5/2} N_{2.5}  
\nonumber \\
&& \quad +O(\epsilon^3) \,, \nonumber \\
K^j &=&  K_1^j + \epsilon^{1/2} K_{1.5}^j +\epsilon K_2^j +\epsilon^{3/2} K_{2.5}^j  +O(\epsilon^{2}) \,, 
\nonumber \\
B &=& B_1 + \epsilon^{1/2} B_{1.5} 
+\epsilon B_2 + \epsilon^{3/2} B_{2.5} 
 +O(\epsilon^2) \,, \nonumber \\
B^{ij} &=& B_2^{ij} + \epsilon^{1/2} B_{2.5}^{ij} 
 +O(\epsilon) \,,
\nonumber \\
K_{\rm ae}^j &=& K_{\rm ae1}^j+ \epsilon^{1/2} K_{\rm ae1.5}^j + \epsilon K_{\rm ae2}^j  +\epsilon^{3/2} K_{\rm ae2.5}^j
\nonumber \\
&& \quad  +O(\epsilon^{2}) \,, 
\nonumber \\ 
R &=& R_1 + \epsilon^{1/2} R_{1.5} + \epsilon R_2 +  \epsilon^{3/2} R_{2.5} +O(\epsilon^{2}) \,,
\nonumber \\ 
X_{K{\rm ae}} &=& X_{K{\rm ae}1} +  \epsilon^{1/2} X_{K{\rm ae}1.5} +  \epsilon X_{K{\rm ae}2}   
\nonumber \\
&& \quad  +  \epsilon^{3/2} X_{K{\rm ae}2.5} +O(\epsilon^{2}) \,,
\nonumber \\ 
X_B &=& X_{B2} + \epsilon^{1/2} X_{B2.5} + O(\epsilon) \,,
\nonumber \\
Y_R &=& Y_{R2} + \epsilon^{1/2} Y_{R2.5} + O(\epsilon) \,,
\nonumber \\
Y_{K{\rm ae}} &=& Y_{K{\rm ae}2} + \epsilon^{1/2} Y_{K{\rm ae}2.5} + O(\epsilon) \,,
\label{expandNKB}
\end{eqnarray}
where the subscript on each term indicates the level (1PN, 2PN, 2.5PN,
etc.) of its leading contribution to the equations of motion, and where we also include the superpotential functions needed to construct the metric.

\subsection{Newtonian, 1PN and 1.5PN solutions}
\label{sec:N1.5PNsolution}

At lowest order in the PN expansion, we only need to evaluate
$\tau^{00} = (-g)T^{00} + O(\rho\epsilon) = \sigma  +
O(\rho\epsilon)$ (recall that $\sigma^{ii} \sim \epsilon \sigma$).
Since the density has compact support, the outer integral vanishes, and we
find
\begin{equation}
N_0 = \frac{8G_0 U}{2-c_{14}} \,.
\label{eq:newtonian}
\end{equation}
The metric to Newtonian order is given by the leading term in Eq.\ (\ref{eq:metric}), $g_{00} = -1+N/2$.
Using Eq.\ (\ref{eq5:GAEtheory}) to relate $G_0$ to $G$, we obtain $N_0 = 4GU$, $g_{00} = -1+2GU$ and $-g = 1+ 4GU + O(\epsilon^2)$.

To the next PN order, we obtain, from Eqs. (\ref{eq:tau}),
(\ref{eq:LambdaT}) and  (\ref{eq:newtonian}),
\begin{eqnarray}
\tau^{00} &=& \sigma - \sigma^{ii} + 4G \sigma U - \frac{7}{8\pi}G \nabla U^2 + O(\rho\epsilon^2) \,, 
\nonumber \\
\tau^{0j} &=& \sigma^j + O(\rho\epsilon^{3/2}) \,, 
\nonumber \\
\tau^{jj} &=& \sigma^{ii} -\frac{1}{8\pi} G\nabla U^2 + 
O(\rho\epsilon^2) \,, 
\nonumber \\
\tau_{\rm ae}^j &=& \sigma_{\rm ae}^j + O(\rho\epsilon^{3/2}) \,.
\label{tauPN}
\end{eqnarray}
Substituting into Eqs.\ (\ref{bigexpansion}), and calculating terms
through 1.5PN order (e.g. $O(\epsilon^{3/2})$ in $N$), we obtain, 
\begin{eqnarray}
N_1 &=&  7G^2U^2-4G\Phi_1+2G^2 \Phi_2+2G{\ddot X} \,,
 \nonumber \\
K_{1}^j &=& 4 \left ( 1-\frac{1}{2} c_{14} \right ) GV^j \,,
\nonumber \\
B_1 &=& \left ( 1-\frac{1}{2} c_{14} \right ) \left [ G^2 U^2+4G\Phi_1-2G^2 \Phi_2 \right ] \,, 
\nonumber \\
K_{\rm ae1}^j &=& - 2  \left ( 1-\frac{1}{2} c_{14} \right ) G c_1^{-1} V_{\rm ae}^j \,,
\nonumber \\
R_1 &=& c_{14} G \dot{X} + 2G \left ( 1-\frac{1}{2} c_{14} \right ) X_{{\rm ae},j}^j \,,
\nonumber \\
X_{K{\rm ae}1} &=&- \frac{2G}{c_1} \left ( 1-\frac{1}{2} c_{14} \right ) X_{{\rm ae},j}^j \,,
\label{eq:1PNsol}
\end{eqnarray}
and
\begin{eqnarray}
N_{1.5} &=& -\frac{2}{3} G\stackrel{(3)\quad}{{\cal I}^{kk}} -\frac{4}{3} G x^k {\ddot {\cal I}}_{\rm ae}^{k}
\,,
\nonumber \\
K_{1.5}^j &=& 0 \,,
\nonumber \\
B_{1.5} &=& -2 G \left ( 1-\frac{1}{2} c_{14} \right )\left [\stackrel{(3)\quad}{{\cal I}^{kk}}  +2 {\ddot {\cal I}}_{\rm ae}^{kk} \right ]\,.
\nonumber \\
K_{\rm ae1.5}^j &=& \frac{2G}{c_1 v_T}  \left ( 1-\frac{1}{2} c_{14} \right )  {\dot {\cal I}}_{\rm ae}^{j} \,,
\nonumber \\
R_{1.5} &=& 0 \,,
\nonumber \\
X_{K{\rm ae}1.5} &=& 0 \,.
\label{eq:1.5PNsol}
\end{eqnarray}
As in the GR case, it is straightforward  to show that the outer integrals
and surface terms give no $\cal R$-independent terms. 

We now use Eq.\ (\ref{eq:transform}) to construct the original fields $\tilde{N}$, $\tilde{B}$ etc., and then Eq. (\ref{eq:metric}) to construct the metric to 1.5PN order.  After applying a gauge transformation,
\begin{equation}
x^{\mu'} = x^\mu + \xi^\mu \,,
\end{equation}
with
\begin{align}
\xi_0 &= \frac{1}{2} \left (1 + \frac{1}{2} c_{14} (3+v_L^{-2} ) \right ) G{\dot X} 
\nonumber \\
& \quad + \frac{1}{2} \left ( 1-\frac{1}{2} c_{14} \right ) (3+v_L^{-2} )G X_{{\rm ae},k}^k 
\nonumber \\
& \quad 
- \frac{2}{3} G \ddot{\cal I}^{kk} - G \dot{\cal I}_{\rm ae}^{kk} - \frac{1}{3} G x^k \dot{\cal I}_{\rm ae}^k \,,
\nonumber \\
\xi_j &= \frac{1}{3} G {\cal I}_{\rm ae}^j \,,
\end{align}
we obtain the 1.5PN metric
\begin{align}
g_{00} &= -1 + 2GU - 2G^2 U^2 \,,
\nonumber \\
g_{0j} &= -4 \left ( 1-\frac{c_{14}}{2} \right ) GV^j - \frac{1}{2} \left ( 1- \frac{c_{14}}{2} (1-v_L^{-2} ) \right ) G\dot{X}_{,j}
\nonumber \\
& \quad + \frac{1}{2}\left ( 1-\frac{c_{14}}{2} \right ) (1-v_L^{-2} ) G X_{{\rm ae},jk}^k \,,
\nonumber \\
g_{jk} &= \delta_{jk} \left ( 1+2GU \right ) \,.
\label{eq:1.5PNmetric}
\end{align}
In the absence of self-gravitating bodies, the source of the {\AE}ther field vanishes, or if the bodies are weakly self-gravitating, the {\AE}ther effects are of one PN order higher; in either case we can set $X^k_{\rm ae} = 0$, and read off the PPN parameters
\begin{align}
\gamma &=1 \,, \quad \beta  = 1 \,,
\nonumber \\
\alpha_1 & = - 4 c_{14} \,, \quad  \alpha_2 = -\frac{1}{2} c_{14} \left (1- \frac{1}{v_L^2} \right ) \,,
\end{align}
with the remaining parameters vanishing (see Eq.\ (\ref{eq5:PPNAE})).  The {\AE}ther field to 1.5PN order is given by
\begin{align}
\tilde{K}^j_{\rm ae} &= - 2  \left ( 1-\frac{1}{2} c_{14} \right ) G c_1^{-1} V_{\rm ae}^j +4 \left ( 1-\frac{1}{2} c_{14} \right ) GV^j 
\nonumber \\
& \quad + \frac{1}{2} W_L G\dot{X}_{,j} + \frac{1}{c_{14}} \left ( 1-\frac{c_{14}}{2} \right ) \left ( W_L - W_T \right ) GX^k_{{\rm ae},jk}
\nonumber \\
& \quad 
+ \frac{2}{c_1 v_T} \left ( 1-\frac{c_{14}}{2} \right ) G\dot{\cal I}_{\rm ae}^j \,.
\end{align}
This is in agreement with standard results \cite{2006PhRvD..73f4015F,tegp2}.  Note that the 1PN solution obtained by Foster and Jacobson \cite{2006PhRvD..73f4015F} made use of specifically tailored gauge choices in order to disentangle the coupled linear field contributions at 1PN order.  It was not clear how to extend that procedure to higher orders; this is in part why we chose to redefine the fields in a manner that could be systematically extended to higher orders.

Notice that there are apparently no 1.5PN radiation reaction terms in the metric.  As in GR, the 1.5PN terms proportional to $\dddot{\cal I}^{kk}$ that appeared in $N_{1.5}$ are pure gauge; but in addition the dipole and monopole {\AE}ther terms $x^k \ddot{\cal I}_{\rm ae}^k$ and $\ddot{\cal I}_{\rm ae}^{kk}$ are also pure gauge.   This does not imply, however, that there is no dipole radiation reaction in this theory; those effects will enter via the modified geodesic equation for compact self-gravitating bodies 
\cite{2022PhRvD.106f4021T}. The same situation occurs in scalar-tensor theory \cite{2013PhRvD..87h4070M}.

\begin{widetext}

\subsection{$B^{jk}$, $K^j$ and $K_{\rm ae}^j$ to 2.5PN order}

Substituting our solutions for the fields to 1.5PN order into Eqs.\ (\ref{eq:tau}) and (\ref{eq:LambdaT}), we obtain
\begin{equation}
\tau^{jk} = \sigma^{jk} + \frac{1}{4\pi} G \left ( U^{,j} U^{,k} - \frac{1}{2} \delta^{jk} |\nabla U|^2 \right ) \,,
\end{equation} 
with the solutions
\begin{align}
B_2^{jk} &= 4G \left ( 1-\frac{c_{14}}{2} \right )  \biggl [  \Phi_1^{jk} + G P_2^{jk}
- \frac{G}{4} \delta^{jk} (2\Phi_2 - U^2 ) \biggr ] \,,
\nonumber \\
B_{2.5}^{jk} &= -2G  \left ( 1-\frac{c_{14}}{2} \right ) \left ( \stackrel{(3)\quad}{{\cal I}^{jk}} + 2 \ddot{\cal I}_{\rm ae}^{(jk)} \right )
\,.
\end{align}

For $K^j$, we substitute the lower-order solutions into 
\begin{equation}
\tau^{0j} = (1+4GU)  \sigma^j + ({16\pi G_0})^{-1} \Lambda_T^{0j} \,,
\end{equation}
and use Eq.\ (\ref{bigexpansionb}) to obtain
\begin{align}
K_2^j &= 8G^2 \left ( 1-\frac{c_{14}}{2} \right ) \biggl [ V_2^j - (1-c_{14}) \Phi_2^j + UV^j 
+ 2(1-c_2)G_7^j +\frac{1}{c_1} \left (c_2 - \frac{c_{14}}{2} \right ) G_{7{\rm ae}}^j
-\frac{1}{4v_L^2} \left ( UV_{\rm ae}^j + \Phi_{2{\rm ae}}^j - V_{2{\rm ae}}^j \right )
\nonumber \\
& \qquad
-2c_2 P(U_{,jk}V^k) + \frac{c_2}{c_1} P(U_{,jk}V_{\rm ae}^k)
+ \frac{c_2}{4c_{14}}(W_L - W_T) \left (UX_{{\rm ae},jk}^k + U_{,j} X_{{\rm ae},k}^k + 2P(U_{,j} V_{{\rm ae},k}^k)\right )
\nonumber \\
& \qquad
-\frac{1}{4} (W_L - W_T)  \left (\Sigma(X_{{\rm ae},jk}^k) -\frac{c_2}{c_{14}}  \Sigma_{,j}(X_{{\rm ae},k}^k) \right )
+ \frac{c_2}{2c_{14}} (3-6c_{14}-2W_L-W_T) P(U V_{{\rm ae},jk}^k) 
 \biggr ]
\nonumber \\
& \quad 
+ G^2 \biggl [2(6+9c_2+c_2 W_L) G_8^j 
+ 4c_2 (6-3c_{14}-W_L) G_9^j +c_{14} W_L \Sigma(\dot{X}_{,j})
+ c_2 W_L \left( U\dot{X}_{,j} + U_{,j} \dot{X} -\Sigma_{,j}(\dot{X}) \right) \biggr ]
\nonumber \\
& \quad
+2 G \left ( 1-\frac{c_{14}}{2} \right ) \ddot{X}^j \,,
\nonumber \\
K_{2.5}^j &= \frac{2}{9} \left ( 1-\frac{c_{14}}{2} \right )G \biggl [ 
 3 x^k \stackrel{(4)}{{\cal I}^{jk}}
- \stackrel{(4)\quad}{{\cal I}^{jkk}}
+2 \epsilon^{mjk}  \stackrel{(3)\quad}{{\cal J}^{mk}}
+6 x^k \stackrel{(3)\quad}{{\cal I}_{\rm ae}^{(jk)}}
- \stackrel{(3)\quad}{{\cal I}_{\rm ae}^{jkk}} 
-2 \stackrel{(3)\quad}{{\cal I}_{\rm ae}^{kjk}} 
+ \frac{18}{c_1v_T} G \left (c_{14} U\dot{\cal I}_{\rm ae}^j +c_2 X_{,jk} \dot{\cal I}_{\rm ae}^k \right ) \biggr ] \,.
\nonumber \\
& \quad 
\end{align}
Finally, for $K_{\rm ae}^j$, we substitute the 1.5PN solutions into
\begin{equation}
\tau_{\rm ae}^j = \sigma_{\rm ae}^j + (8\pi G_0)^{-1} \Lambda_{\rm ae}^j \,,
\end{equation}
and use Eq.\ (\ref{bigexpansiond}) to obtain
\begin{align}
K_{\rm ae 2}^j &=-\frac{2G^2}{c_1}  \left ( 1-\frac{c_{14}}{2} \right ) \biggl [ c_{14} V_2^j +c_{14} \Phi_2^j + (6c_1-c_{14}) UV^j 
+ 2(3c_1-2c_2)G_7^j +\frac{1}{c_1} \left (2c_2 +c_1 - c_{14} \right ) G_{7{\rm ae}}^j
\nonumber \\
& \qquad
-  UV_{\rm ae}^j - 2 \Phi_{2{\rm ae}}^j + 3 V_{2{\rm ae}}^j 
+(3c_1-c_{14}- 2c_2) \left ( 2P(U_{,jk}V^k) - \frac{1}{c_1} P(U_{,jk}V_{\rm ae}^k) \right )
\nonumber \\
& \qquad
+\frac{1}{2} (W_L - W_T)  \left (\Sigma(X_{{\rm ae},jk}^k) +\frac{3c_1-c_{14}-2c_2}{2c_{14}}  \Sigma_{,j}(X_{{\rm ae},k}^k) \right )
\nonumber \\
& \qquad
+ \frac{1}{2c_{14}} \left (12 c_{14}c_1 + (3c_1 -c_{14} + 2c_2 )(W_L - W_T)\right) P(U V_{{\rm ae},jk}^k) 
\nonumber \\
& \qquad
- \frac{1}{2c_{14}} \left (\frac{2c_{14}}{c_2}(3c_1 -c_{14} ) + (3c_1 -c_{14}-2c_2 )(W_L - W_T)\right) P(U_{,j} V_{{\rm ae},k}^k)
\nonumber \\
& \qquad
+ \frac{1}{4c_{14}} (W_L - W_T) \left ( (3c_1 -c_{14} + 2c_2 ) U X_{{\rm ae},jk}^k - (3c_1 -c_{14} - 2c_2 ) U_{,j} X_{{\rm ae},k}^k \right )
 \biggr ]
\nonumber \\
& \quad 
- \frac{G^2}{4c_1} \biggl [ \left ( 8(2-c_{14})(3c_2+c_{14}-3c_1 )+2(3c_1-c_{14}-4c_2 )W_L \right )G_8^j 
\nonumber \\
& \qquad 
+  \left ( 36c_2 +12c_{14} -24c_1 (2-c_{14}) + 2(3c_1-c_{14}+2c_2 )W_L \right )G_9^j 
\nonumber \\
& \qquad
 + (3c_1 -c_{14} + 2c_2 )W_L   U \dot{X}_{,j}
- (3c_1 -c_{14} - 2c_2 )W_L \left(  U_{,j} \dot{X} -\Sigma_{,j}(\dot{X}) \right)  +2c_{14} W_L \Sigma(\dot{X}_{,j})
\biggr ]
\nonumber \\
& \quad
- \frac{G}{c_1} \left ( 1-\frac{c_{14}}{2} \right )(1-W_T) \ddot{X}_{\rm ae}^j \,,
\nonumber \\
K_{{\rm ae}2.5}^j &=\frac{G}{c_1}  \left ( 1-\frac{c_{14}}{2} \right ) \biggl [\frac{1}{3v_T^3}  \left ( r^2 \stackrel{(3)}{{\cal I}_{\rm ae}^j}
-2x^k \stackrel{(3)}{{\cal I}^{jk}_{\rm ae}}
+ \stackrel{(3)\quad}{{\cal I}^{jkk}_{\rm ae}} \right )
- \frac{1}{c_1v_T} G \left ( 6 c_1 U\dot{\cal I}_{\rm ae}^j -(3c_1 -c_{14} - 2c_2 ) X_{,jk} \dot{\cal I}_{\rm ae}^k \right ) \biggr ] \,.
\nonumber \\
& \quad 
\end{align}

\subsection{$N$ and $B$ to 2.5PN order}

Using the 1.5PN metric (prior to the gauge transformation to the PPN gauge), we find that, to the required order,
\begin{align}
-g &= 1 + 4  GU + \biggl [\left (1-\frac{c_{14}}{2} \right ) ( 4 G^2 \Phi_2 -8 G\Phi_1 )+ (6+c_{14}) G^2 U^2 + \left ( 2-3c_{14}+\frac{c_{14}}{v_L^2} \right ) G \ddot{X} 
\nonumber \\
& \qquad -2 \left (1-\frac{c_{14}}{2} \right ) \left ( 3 - \frac{1}{v_L^2} \right )G \dot{X}_{{\rm ae},k}^k \biggr ]
+  \biggl [ \frac{2}{3} (2-3c_{14}) G \stackrel{(3)\quad}{{\cal I}^{kk}} + 4 (1-c_{14} ) G \ddot{\cal I}_{\rm ae}^{kk} - \frac{4}{3} G x^k  \ddot{\cal I}_{\rm ae}^{k} \biggr ]+ O(\epsilon^3) \,.
\end{align}
Inserting this along with the 1.5PN solutions for the fields into the expressions for $\tau^{00}$ and $\tau^{jj}$ in Eqs.\ (\ref{eq:tau}) and (\ref{eq:LambdaT}), we obtain
\begin{align}
N_2 &=  4 \left (1-\frac{c_{14}}{2} \right ) G^2 \biggl [2GU\Phi_2 - 4 U\Phi_1 - (1+2c_{14}) V^jV^j  - 4 \Sigma(\Phi_1)  + 2 \Sigma_j (V^j) \biggr ] -2 G \ddot{X}_1 + G^2 \ddot{X}_2 + \frac{1}{6} G \stackrel{(4)}{Y}
\nonumber \\
& \quad + G^2 \biggl [ \left (7+4c_{14}- \frac{10c_{14}}{2-c_{14}} W_L \right ) U\ddot{X}
+  \left (1-12c_{14}- \frac{c_{14}}{2-c_{14}} W_L \right ) \Sigma(\ddot{X}) +\frac{2}{3} G (10+3c_{14}) U^3  
\nonumber \\
& \qquad
-4\left (1+7c_{14}- \frac{c_{14}}{2-c_{14}} W_L \right ) G_1
-8\left (2+3c_{14}- \frac{c_{14}}{2-c_{14}} W_L \right ) G_2
+8 (4-5c_{14}) G_3 
\nonumber \\
& \qquad
+24 \left (1-\frac{c_{14}}{2} \right ) G_4
-16 (1+c_{14}) G_5 - 16 \left (1-\frac{c_{14}}{2} \right ) G_6 -16 \left (1-\frac{c_{14}}{2} \right ) GH \biggr ]
\nonumber \\
& \quad
+ c_{14} G^2 \biggl [ \left (4+ \frac{3}{2-c_{14}}W_L \right ) \left ( \dot{\Sigma}(\dot{X}) -\dot{U}\dot{X} \right )
- 8\Sigma^{j} (\dot{X}_{,j}) + 4 (2-c_{14}-W_L) V^j \dot{X}_{,j} 
-\left ( c_{14}+ \frac{W_L^2}{4-2c_{14}} \right ) \dot{X}_{,j}\dot{X}_{,j}
\nonumber \\
& \qquad
+ \frac{4}{c_1} \left \{ 3 G_{3{\rm ae}} 
+3G_{5{\rm ae}}  
+ \frac{1}{2} W_L  \dot{X}_{,j} V_{\rm ae}^j 
+ 2 \left (1-\frac{c_{14}}{2} \right ) \left ( V^j V_{\rm ae}^j+\Sigma^{j}(V_{\rm ae}^j) -  \Sigma_{\rm ae}^j(V^j) -2G_{4{\rm ae}} \right ) 
\right \}\biggr ]
\nonumber \\
& \quad
- \frac{2}{c_1}  \left (1-\frac{c_{14}}{2} \right )G^2 \biggl [ (2+ 4W_T) G^{\rm ae}_{4{\rm ae}}
+ 3 V_{\rm ae}^j V_{\rm ae}^j - 2 (W_L-W_T) V_{\rm ae}^j X^k_{{\rm ae},jk}
-6 \Sigma_{\rm ae}^{j}(V_{\rm ae}^j) + 4(1+W_T) P(V_{\rm ae}^j V_{{\rm ae},jk}^k ) \biggr ]
\nonumber \\
& \quad
+ \left (1-\frac{c_{14}}{2} \right )G^2 \biggl [ 
2 \left (4-\frac{10W_L-3W_T}{2-c_{14}} \right ) U \dot{X}_{{\rm ae},k}^{,k}
-2 \left (4+\frac{3(W_L-W_T)}{2-c_{14}} \right ) \dot{U} {X}_{{\rm ae},k}^{,k}
-16 \Sigma^j (X^k_{{\rm ae},jk})
\nonumber \\
& \qquad
- 2\left (2c_{14}+\frac{W_L(W_L-W_T)}{2-c_{14}} \right ) \dot{X}_{,j} X_{{\rm ae},jk}^k
- \frac{2}{c_{14}}(1+W_T)(W_L-W_T) X_{{\rm ae},k}^k V^m_{{\rm ae},m}
- 2\left (12+ \frac{W_L}{2-c_{14}} \right )\Sigma(\dot{X}^j_{{\rm ae},j})
\nonumber \\
& \qquad
+8(2-c_{14} -W_L+W_T) V^j X^k_{{\rm ae},jk}
+2 \left (4+\frac{3(W_L-W_T)}{2-c_{14}} \right )\dot{\Sigma}(X^j_{{\rm ae},j})
+16 (1+W_T)P(V^j V^k_{{\rm ae},jk})
\nonumber \\
& \qquad
-\frac{2}{c_{14}} (W_L-W_T) \left ( (3-W_T)\Sigma^j_{\rm ae}(X^k_{{\rm ae},jk}) 
- (1+W_T)\Sigma^j_{{\rm ae},j}(X^k_{{\rm ae},k})  \right )
\biggr ]
 -G^2  W_L(1+W_T) \left (\dot{X} V^j_{{\rm ae},j} - \Sigma^j_{{\rm ae},j}(\dot{X}) \right )
\nonumber \\
& \quad
- G^2 W_L (3-W_T) \Sigma^j_{\rm ae} (\dot{X}_{,j})
-  2\left (1-\frac{c_{14}}{2} \right )^2 G^2 \left (2 + \frac{(W_L-W_T)^2}{c_{14}(2-c_{14})} \right ) X^k_{{\rm ae},jk}X^m_{{\rm ae},jm}
\nonumber \\
& \quad
-2G^2 (6-12 c_{14}-4W_L-3W_T) P(U\dot{V}^j_{{\rm ae},j})
-\frac{4}{c_{14}} \left (1-\frac{c_{14}}{2} \right ) G^2 [1-2c_{14}-W_T(1+W_T-W_L)] P(V^j_{{\rm ae},j}V^k_{{\rm ae},k})
\nonumber \\
& \quad
-2G^2 [14-16c_{14}-3W_T-W_L(2-W_T)] P(\dot{U}V^j_{{\rm ae},j}) \,,
\nonumber \\
B_2 &=  4\left (1-\frac{c_{14}}{2} \right )^2 G^2 \biggl [V^j V^j -2 \Sigma^j(V^j)+ 2G_4 \biggr ]
+\left (1-\frac{c_{14}}{2} \right ) G^2 \biggl [16 \Sigma^{jj}(U) - \ddot{X}_2  -8(6c_2-c_{14})G_3 +16(1-3c_2)G_5 \biggr ]
\nonumber \\
& \quad
+G^2 \biggl [ \left ( 1+\frac{5}{2}c_{14}-3c_2(1-2c_{14}) \right )U\ddot{X}
- \left ( 1-\frac{c_{14}}{2} (1+W_L) \right ) \Sigma(\ddot{X}) 
\biggr ]
+ 2 G \left (1-\frac{c_{14}}{2} \right ) \ddot{X}_1
\nonumber \\
& \quad
-2 \left (10 +3c_{14}^2 +6c_2(5-c_{14})-c_{14}(5-W_L) \right ) G^2 G_1
- 12 \left (c_{14}+ 5c_2 - c_{14}c_2 \right ) G^2 G_2
\nonumber \\
& \quad
+\frac{1}{2} G^2 (6c_2 - c_{14})W_L  \left ( \dot{\Sigma}(\dot{X}) - \dot{U} \dot{X} \right )
- \frac{1}{c_1}\left (1-\frac{c_{14}}{2} \right ) \left ( 6c_2-c_{14}-5c_1-12c_1 c_2 \right ) G^2 U \dot{X}^j_{{\rm ae},j}
\nonumber \\
& \quad
+ \frac{4}{c_1} \left (1-\frac{c_{14}}{2} \right ) G^2 \biggl [ (6c_2 - c_{14}) \left ( G_{3{\rm ae}} +G_{5{\rm ae}} \right ) - \left (1-\frac{c_{14}}{2} \right )  \left (G^{\rm ae}_{4{\rm ae}} 
+ \frac{1}{2}V_{\rm ae}^j V_{\rm ae}^j \right )
 \biggr ]
+ \left (1-\frac{c_{14}}{2} \right ) W_L G^2  \dot{X} V^j_{{\rm ae},j}
\nonumber \\
& \quad
- \frac{1}{c_{14}} \left (1-\frac{c_{14}}{2} \right ) ( 6c_2-c_{14})(W_L-W_T) G^2 \left (\dot{U} {X}^j_{{\rm ae},j} -\dot{\Sigma}( {X}^j_{{\rm ae},j}) \right )
+ \left (1-\frac{c_{14}}{2} \right ) W_L G^2 \Sigma(\dot{X}^j_{{\rm ae},j})
\nonumber \\
& \quad
+ \frac{2}{c_{14}}  \left (1-\frac{c_{14}}{2} \right )^2  (W_L-W_T) G^2 \biggl [ X^j_{{\rm ae},j}V^j_{{\rm ae},j}
+ \Sigma^j_{\rm ae}({X}^k_{{\rm ae},jk}) -  \Sigma^j_{\rm ae}({X}^k_{{\rm ae},k})_{,j} \biggr ]
+ \frac{4}{c_1}\left (1-\frac{c_{14}}{2} \right )^2 G^2  \Sigma_{\rm ae}^j(V^j_{\rm ae}) 
\nonumber \\
& \quad
+ \left (1-\frac{c_{14}}{2} \right ) W_L G^2 \left ( \Sigma^j_{\rm ae}(\dot{X}_{,j}) - \Sigma^j_{\rm ae}(\dot{X})_{,j} \right ) +\frac{8}{c_1} \left (1-\frac{c_{14}}{2} \right )^2 G^2 \left [ P(V_{\rm ae}^j V^k_{{\rm ae},jk}) - 2c_1 P(V^j V^k_{{\rm ae},jk}) \right ]
\nonumber \\
& \quad
+2 \left (1-\frac{c_{14}}{2} \right ) \left ( 6-12c_{14} - 4W_L-W_T- \frac{6c_2}{c_{14}} (W_L-W_T) \right ) G^2  P(\dot{U}V^j_{{\rm ae},j}) 
\nonumber \\
& \quad
+ \frac{2}{c_1}  \left (1-\frac{c_{14}}{2} \right )G^2 ((12c_2-13)c_1-6c_2+c_{14}) P(U\dot{V}^j_{{\rm ae},j})
\nonumber \\
& \quad
+\frac{4}{c_{14}} \left (1-\frac{c_{14}}{2} \right )^2 G^2 \left ( 3 - 6c_{14}-2W_L-W_T   \right ) P(V^j_{{\rm ae},j}V^k_{{\rm ae},k})
\,,
\nonumber \\ 
R_2 &= - c_{14} \biggl [ 7G^2 P(U\dot{U})+ G\dot{X}_1- \frac{1}{2} G^2 \dot{X}_2 - \frac{1}{12}G \dddot{Y} \biggr ] - c_1 X_{K{\rm ae}2} \,,
\nonumber \\
c_1 X_{K{\rm ae}2} &= 2  \left (1-\frac{c_{14}}{2} \right ) G^2 \biggl [ 8c_2 P(U_{,j}V^j) 
 - 4\frac{c_2}{c_{1}} P(U_{,j} V_{\rm ae}^j)
- \frac{c_2}{c_{14}} (W_L-W_T) U X^j_{{\rm ae},j} 
+\frac{1}{2} W_T X^j_{\rm ae}(U_{,j}) 
+ \frac{1}{2} W_T X(V^j_{{\rm ae},j}) 
\nonumber \\
& \qquad
-\frac{1}{2}(4-W_T) X^j_{\rm ae}(U)_{,j} 
-\frac{1}{2} (W_L-W_T) X(X^k_{{\rm ae},jk})_{,j}
-2c_{14} X(V^j)_{,j} + \frac{c_{14}}{c_1} X(V_{\rm ae}^j)_{,j} 
\nonumber \\
& \qquad
+ \frac{1}{c_1} (4c_1c_2+3c_1-c_{14}-2c_2) P(UV^j_{{\rm ae},j})
+ \frac{c_2}{c_{14}}(W_L-W_T) \Sigma(X^j_{{\rm ae},j}) \biggr ]
- \frac{1}{6 v_T^2} \left (1-\frac{c_{14}}{2} \right ) G \ddot{Y}^j_{{\rm ae},j}
\nonumber \\
& \quad
+c_2 W_L G^2  \left (\Sigma(\dot{X}) - U \dot{X} \right ) 
- \frac{1}{2} c_{14} W_L G^2 X(\dot{X}_{,j})_{,j}
+ 4\left ((c_{14}-5)c_2 - c_{14} \right ) G^2 P(U\dot{U})
 \,,
\nonumber \\
X_{B2} &= - 2 \left (1-\frac{c_{14}}{2} \right ) \biggl [ G^2 P(U^2) - 2GX_1 + G^2 X_2 \biggr ] \,,
\nonumber \\
Y_{R2} & = c_{14} G \dot{Y} + 2G  \left (1-\frac{c_{14}}{2} \right ) Y^j_{{\rm ae},j} \,,
\nonumber \\
Y_{K{\rm ae}2} &= Y^j_{{\rm ae},j} \,,
\end{align}
and
\begin{align}
N_{2.5} &= - \frac{1}{30}  G\biggl \{ (4x^{kl}+2r^2\delta^{kl})
\stackrel{(5)\quad}{{\cal I}^{kl}}
- 4 x^k \stackrel{(5)\quad}{{\cal I}^{kll}}
+ \stackrel{(5)\quad}{{\cal I}^{kkll}} \biggr \}
+ \frac{16}{3} \left (1-\frac{3c_{14}}{2} \right )G^2 U \stackrel{(3)\quad}{{\cal I}^{kk}}
\nonumber \\
& \quad
-4 \left (1-\frac{c_{14}}{2} \right )G^2 \left (\stackrel{(3)\quad}{{\cal I}^{jk}} + 2 \ddot{\cal I}^{(jk)}_{\rm ae} \right ) X_{,jk} +16 (1-c_{14}) G^2 U \ddot{\cal I}^{kk}_{\rm ae}
-\frac{2}{15}  Gr^2 x^k  \stackrel{(4)\quad}{{\cal I}_{\rm ae}^{k}}
- \frac{6}{v_T^3} \dot{\cal I}^j_{\rm ae} G^2 \dot{X}_{,j}
\nonumber \\
& \quad
- \frac{8}{v_T^3} \left (1-\frac{c_{14}}{2} \right )G^2 \dot{\cal I}^j_{\rm ae}  \left (2V^j + \frac{1}{2c_{14}} (1+W_T) X^k_{{\rm ae},jk} \right )
 -\frac{16}{3} G^2 \ddot{\cal I}^j_{\rm ae} \Sigma(x^j) 
-\frac{2}{3} \left ( 7 + \frac{9}{v_T^3} \right ) G^2  \ddot{\cal I}^j_{\rm ae} X_{,j} \,,
\nonumber \\
B_{2.5} &= - \frac{1}{9}  \left (1-\frac{c_{14}}{2} \right ) G  \biggl [ 
3 r^2 \stackrel{(5)\quad}{{\cal I}^{kk}}
-2x^l \stackrel{(5)\quad}{{\cal I}^{kkl}}
- 8 x^l \epsilon^{mlk} \stackrel{(4)\quad}{{\cal J}^{mk}}
+ 6 \stackrel{(3)\quad}{M^{kkll}}
 +6r^2 \stackrel{(4)\quad}{{\cal I}_{\rm ae}^{kk}}
+6x^l \stackrel{(4)\qquad}{{\cal I}_{\rm ae}^{(kkl)}} \biggr ]
\nonumber \\
& \quad
+  \left (1-\frac{c_{14}}{2} \right ) G^2  \biggl [ 
\frac{2}{c_1 v_T} \left ( (c_{14}-6c_2) \left ( \dot{\cal I}^j_{\rm ae} \dot{X}_{,j} +  \ddot{\cal I}^j_{\rm ae} {X}_{,j} \right ) + (2-c_{14}) \dot{\cal I}^j_{\rm ae} X^k_{{\rm ae},jk} \right )
- \frac{2}{3} \ddot{\cal I}^j_{\rm ae} {X}_{,j} \biggr ] \,,
\nonumber \\
R_{2.5} & = - \frac{1}{18} c_{14} G r^2 \biggl [  \stackrel{(4)\quad}{{\cal I}^{kk}} 
+ \frac{6}{5}  x^j \stackrel{(3)}{{\cal I}^{j}_{\rm ae}} \biggr ]  - c_1 X_{K{\rm ae}2.5} \,,
\nonumber \\
c_1 X_{K{\rm ae}2.5} &=  - \frac{2}{9v_T}  \left (1-\frac{c_{14}}{2} \right ) G \biggl [ c_{14} r^2 
\left ( \stackrel{(3)\quad}{{\cal I}_{\rm ae}^{kk}} - \frac{3}{5} x^j  \stackrel{(3)\quad}{{\cal I}_{\rm ae}^{j}} \right ) - 9 G(c_{14}+2c_2 )  \dot{\cal I}_{\rm ae}^{j} X_{,j} \biggr ] \,,
\nonumber \\
X_{B2.5} & = -\frac{2}{3} \left (1-\frac{c_{14}}{2} \right ) G r^2 \left [ \stackrel{(3)\quad}{{\cal I}^{kk}}  +2 {\ddot {\cal I}}_{\rm ae}^{kk} \right ] \,,
\end{align}
\end{widetext}
with $Y_{R2.5} = Y_{K{\rm ae}2.5} = 0$.  It is then straightforward to construct the physical metric and the {\AE}ther field using Eqs.\ (\ref{eq:metric}) and (\ref{eq:transform}).

\section{Future prospects and concluding remarks}
\label{sec:conclusions}

We have applied post-Minkowskian theory to the \eae theory, and demonstrated that, after a field transformation, the relaxed field equations can be put into a form that parallels that of general relativity, and that is suitable for obtaining solutions to high orders in a post-Newtonian expansion.  As an application of the method, we obtained explicit solutions for the fields through 2.5PN order, in terms of Poisson-like potentials and superpotentials constructed from the matter densities.  

In a forthcoming publication we will use these results to obtain the equations of motion for compact binaries through 2.5PN order.  We will use the prescription pioneered by Eardley \cite{1975ApJ...196L..59E} for treating gravitationally bound bodies in alternative theories of gravity, in which one assumes that each body's mass is a function of an invariant quantity constructed from the auxiliary field(s) of the theory, evaluated at the location of the body.  For scalar-tensor theory (Eardley's original motivation) it is the scalar field itself; for \eae theory, the conventional choice is the invariant $\gamma \equiv - \tilde{K}^\mu u_\mu$, where $u^\mu$ is the four-velocity of the body (the other possible invariant $\tilde{K}^\mu \tilde{K}_\mu $ is unity by definition, and thus trivial).  
This results in a modified geodesic equation for each body, given by (see, eg.\ \cite{2022PhRvD.106f4021T})
 \begin{align}
u_A^\nu \nabla_\nu &\left [ m_A u_{A\alpha} + m'_A \tilde{K}^\mu \left (g_{\mu\alpha} + u_{A\mu} u_{A\alpha} 
\right ) \right ] 
\nonumber \\
&\qquad
= m'_A u_{A\mu} \nabla_\alpha \tilde{K}^\mu  \,,
\label{eq:case1}
\end{align}
where $m_A = m_A(\gamma)$, $m'_A \equiv dm_A/d\gamma$.  This paper provides the ingredients needed to obtain the equations of motion to 2.5PN order.

\acknowledgments

This work was supported in part by the National Science Foundation,
Grants No.\ PHY 19-09247 and PHY 22-07681.   We are grateful for the hospitality of  the Institut d'Astrophysique de Paris where part of this work was carried out.


\appendix

\section{Wavelike solutions to the linearized vacuum equations}
\label{app:waves}

Here we analyze the far-zone waves implied by the linearized equations (\ref{eq:linear1}) and (\ref{eq:linear2}) using an extension of the method described in Sec.\ 11.1 of \cite{PW2014}  for decomposing waves in the far-away wave zone in general relativity.   Far from the source we express each field in the generic form
\begin{equation}
A = R^{-1}  A_0 (\tau, \bm{n} ) + O(R^{-2}) \,,
\label{app:1}
\end{equation}
where $\tau = t - R/v_g$, and $\bm{n} = \bm{\nabla} R$.   Then 
\begin{align}
A_{,j} &= - n^j \dot{A}_0/v_g R + O(R^{-2}) \,,
\nonumber \\
\Box A &= - \left (1 - v_g^{-2} \right ) \ddot{A}_0 / R + O(R^{-2}) \,,
\label{app:2}
\end{align}
where a dot denotes $d/d\tau$.  We also decompose the various vector and tensor amplitudes into their irreducible pieces (see, eg. Box 5.7 of  \cite{PW2014}),
\begin{align}
K_0^j &= K_0 n^j + K_{\rm T}^j \,,
\nonumber \\
K_{\rm ae0}^j &= K_{\rm ae0} n^j + K_{\rm aeT}^j \,,
\nonumber \\
B_0^{jk}  &= \frac{1}{3}\delta^{jk}  B_0 + \left ( n^j n^k -  \frac{1}{3} \delta^{jk}  \right ) B_{\rm LTF} 
\nonumber \\
& \qquad + 2 n^{(j} B_{\rm T}^{k)} + B^{jk}_{\rm TT} \,,
\label{app:3}
\end{align}
where the subscripts denote the transverse (T), longitudinal tracefree (LTF) and transverse traceless (TT) parts.  Imposing harmonic gauge $\dot{N} + K^j_{,j} = 0$, and $\dot{K}^j + B^{jk}_{,k} =0$, keeping the leading $1/R$ amplitudes, and decomposing into irreducible parts leads to the four conditions
\begin{align}
K_0 &= v_g N_0 \,,
\nonumber \\
v_g K_0 &= \frac{1}{3} B_0 + \frac{2}{3} B_{\rm LTF} \,,
\nonumber \\
v_g K^j_{\rm T} &= B^j_{\rm T} \,,
\label{eq:gaugeconditions}
\end{align}
where henceforth, we drop the dots.  Under a gauge transformation $x^\alpha \to x^\alpha + \zeta^\alpha$ with
\begin{align}
\zeta^0 &= R^{-1} \alpha (\tau, \bm{n} ) + O(R^{-2}) \,,
\nonumber \\
\zeta^j &= R^{-1} \left [ \beta (\tau, \bm{n} ) n^j + \beta_{\rm T}^j (\tau, \bm{n} ) \right ] + O(R^{-2}) \,,
\end{align}
the amplitudes undergo the changes
\begin{align}
N_0 &\to N_0 + \dot{\alpha} + v_g^{-1} \dot{\beta} \,,
\nonumber \\
K_0 &\to K_0 + v_g^{-1} \dot{\alpha} +  \dot{\beta} \,,
\nonumber \\
K_{\rm T}^j &\to K_{\rm T}^j + \dot{\beta}_{\rm T}^j \,,
\nonumber \\
B_0 &\to B_0 + 3 \dot{\alpha}  - v_g^{-1} \dot{\beta} \,,
\nonumber \\
B_{\rm LTF} &\to B_{\rm LTF} + 2v_g^{-1}  \dot{\beta} \,,
\nonumber \\
B_{\rm T}^j &\to B_{\rm T}^j  + v_g^{-1}  \dot{\beta}_{\rm T}^j  \,,
\nonumber \\
B_{\rm TT}^{jk}  &\to B_{\rm TT}^{jk} \,,
\nonumber \\
K_{\rm ae0} &\to K_{\rm ae0} + \dot{\beta} \,,
\nonumber \\
K_{\rm aeT}^j &\to K_{\rm aeT}^j + \dot{\beta}_{\rm T}^j \,.
\label{eq:gaugechanges}
\end{align}
The time component of the {\AE}ther field, $K_{\rm ae}^0$ is gauge invariant to linear order.

Substituting Eqs.\ (\ref{app:1}) - (\ref{app:3}) (but not the harmonic gauge conditions)
 into Eqs.\ (\ref{eq:linear1}) and (\ref{eq:linear2}) and decomposing into irreducible parts, we obtain the system of nine equations:
 \begin{subequations}
 \begin{align}
&(1-v_g^2) N_0 = \frac{c_{14}}{2}  \left [ N_0 + B_0 + 4v_g (K_{\rm ae0} - K_0) \right ] \,,
\label{eq:wave1}
 \\
&(1-v_g^2) K_0 = \frac{c_{14}}{2} v_g  \left [ N_0 + B_0 + 4v_g (K_{\rm ae0} - K_0) \right ] \,,
\label{eq:wave2}
 \\
&(1-v_g^2) K_{\rm T}^j = 2(c_{14} v_g^2 -c_1 ) (K_{\rm aeT}^j - K_{\rm T}^j ) \,,
\label{eq:wave3}
 \\
&(1-v_g^2) B_0 = -\frac{3}{2} c_2 \left [ v_g^2 (3N_0 -B_0) - 4v_g K_{\rm ae0} \right ] \,,
\label{eq:wave4}
 \\
&(1-v_g^2) B_{\rm LTF}  = 0 \,,
\label{eq:wave5}
 \\
&(1-v_g^2) B_{\rm T}^j =0 \,,
\label{eq:wave6}
 \\
&(1-v_g^2) B_{\rm TT}^{jk}  = 0 \,,
\label{eq:wave7}
 \\
&(c_1-v_g^2 c_{14}) ( K_{\rm aeT}^j - K_{\rm T}^j ) = 0 \,,
\label{eq:wave8}
 \\
&c_{14} v_g  \left [ N_0 + B_0 + 4v_g (K_{\rm ae0} - K_0) \right ] 
\nonumber \\
 &\qquad \qquad = - c_2  \left [ v_g (3N_0 -B_0) - 4 K_{\rm ae0} \right ] \,.
\label{eq:wave9}
\end{align} 
\label{eq:wave10}
\end{subequations}
It is straightforward to show that this system has three distinct eigenvalues for $v_g^2$.

{\em Case 1: $v_g = 1$.}  In this case, $B_{\rm LTF}$, $B_{\rm T}^j $ and $B_{\rm TT}^{jk} $ are unconstrained, and $K_{\rm aeT}^j - K_{\rm T}^j = 0$ (unless $c_4 = 0$).  Combining the scalar parts of the gauge conditions (\ref{eq:gaugeconditions}) and the wave equations (\ref{eq:wave10}) we find that $N_0 = K_0 = (B_0 + 2B_{\rm LTF})/3$ and $K_{\rm ae0} = (3N_0 - B_0)/4$ and thus that $2 K_{\rm ae0} - B_{\rm LTF} = 0$  We can then choose $\alpha$ and $\beta$ so that $N_0$,  $K_0$ and $B_0$ all vanish, and thus so that $K_{\rm ae0}$ and $B_{\rm LTF}$ vanish.   
Also we have that $K_{\rm T}^j = K_{\rm aeT}^j = B_{\rm T}^j$; we can choose $\beta_{\rm T}^j$ to make them all vanish.  In the end, only the gauge invariant $B_{\rm TT}^{jk}$ is unconstrained.  This is a pure transverse traceless metric gravitational wave, with speed unity.  It was the observational constraint on the speed of gravitational waves set by the event GW170817 and GRB170817 that led us to impose the constraint $c_1 + c_3 =0$ in the first place.

{\em Case 2: $v_g = (c_1/c_{14})^{1/2} = v_T$.}  In this case, $K_{\rm aeT}^j $ is unconstrained, while $B_{\rm LTF} = B_{\rm T}^j = B_{\rm TT}^{jk} = K_{\rm T}^j  = 0$.  Examining the four scalar wave equations (\ref{eq:wave1}), (\ref{eq:wave2}), (\ref{eq:wave4}) and (\ref{eq:wave9}), we observe that the determinant of the linear system does not vanish, so that $N_0 = K_0 = B_0 = K_{\rm ae0} = 0$.  This is a pure transverse vector wave, with no metric perturbation, to linear order.

{\em Case 3:}  For this final case, we must consider the five non-transverse scalar wave equations (\ref{eq:wave1}), (\ref{eq:wave2}), (\ref{eq:wave4}), (\ref{eq:wave5})  and (\ref{eq:wave9}).  Requiring the determinant of this system to vanish yields $v_g =1$ (Case 1) plus a solution with speed $v_g=v_L$ given by
\begin{equation}
 v_L^2 =  \frac{c_2 ( 2- c_{14})}{c_{14} (2+ 3c_2)} \,.
\end{equation}
The solutions are $B_{\rm LTF} = B_{\rm T}^j = B_{\rm TT}^{jk} = K_{\rm T}^j  = K_{\rm aeT}^j  = 0$, along with
\begin{align}
N_0 &= \frac{4c_{14}}{2-c_{14}} \frac{v_L}{1-v_L^2} K_{\rm ae0} \,,
\nonumber \\
K_0 &=  v_L N_0 \,,
\nonumber \\
B_0 &= 3 v_L K_0 \,,
\end{align}
with $K_{\rm ae0}$ the unconstrained amplitude.   This is a longitudinal {\AE}ther wave with accompanying longitudinal metric perturbations.

\section{Transformation to new variables}
\label{app:transform}

In this Appendix, we derive the transformation (\ref{eq:transform}) that eliminates all terms linear in the fields $\tilde{N}$, $\tilde{K}^j$, $\tilde{B}^{jk}$ and $\tilde{K}_{\rm ae}^j$ apart from terms that consist of a leading d'Alembertian of the fields.   Those linear terms are displayed in Eqs.\ (\ref{eq:linear1}) and (\ref{eq:linear2}).   It is known from earlier work on \eae theory that the d'Alembertian of $N$ appears in the combination $(1-c_{14}/2) \Box N$, so that the coupling constant $G_0$ is renormalized by that prefactor.  That will be a constraint on the solution.  From the structure of  Eqs.\ (\ref{eq:linear1}) and (\ref{eq:linear2}) it is clear that the combination $\tilde{K}_{\rm ae}^j - \tilde{K}^j$ is prevalent, so we will define $\tilde{K}_{\rm ae}^j = K_{\rm ae}^j + K^j + \dots$ .  We want to remove all offending linear terms in the field equations through 2PN order.  Finally, we will want to investigate the forms taken by the harmonic gauge conditions $\tilde{N}_{,0} + \tilde{K}^j_{,j} =0$ and $\tilde{K}^j_{,0} + \tilde{B}^{jk}_{,k} = 0$ in the new variables.  Because the transformation of $\tilde{N}$ will go through 2PN order, or to relative order $\epsilon^2$, we will want to include terms at relative order  $\epsilon^2$ in the transformation of $\tilde{K}^j$, even though that is a PN order higher in $\tilde{K}^j$ than we actually need for the equations of motion; for completeness, we will also transform $\tilde{K}_{\rm ae}^j$ to the same relative order.
  The second gauge condition does not impose additional conditions on the transformations.

 Accordingly we try a linear transformation of the form:
\begin{align}
\tilde{N} &= N + \epsilon \bigl ( a_2 B + a_3 \ddot{X}_N + a_4 \dot{X}_K + a_5 \dot{X}_{K{\rm ae}} \bigr )
\nonumber \\
& \quad
+ \epsilon^2 \bigl  (  a_6 \ddot{X}_B + a_7 \stackrel{(4)}{{Y}_N} + a_8 \stackrel{(3)}{{Y}_K}
+ a_9 \stackrel{(3)\quad}{{Y}_{K{\rm ae}}} \bigr ) \,,
\nonumber \\
\tilde{B}^{jk} &= {B}^{jk} + \delta^{jk} \biggl [ b_3 \ddot{X}_N + b_4 \dot{X}_K + b_5 \dot{X}_{K{\rm ae}}
\nonumber \\
& \quad + \epsilon \bigl ( b_6 \ddot{X}_B + b_7 \stackrel{(4)}{{Y}_N} + b_8 \stackrel{(3)}{{Y}_K}
+b_9 \stackrel{(3)\quad}{{Y}_{K{\rm ae}}} \bigr ) \biggr ] \,,
\nonumber \\
\tilde{K}^j &= K^j +d_3 \dot{X}_{N,j} + d_4 {X}_{K,j} + d_5 {X}_{{K{\rm ae}},j}
\nonumber \\
& \quad+ \epsilon \bigl ( d_6 \dot{X}_{B,j} + d_7 \stackrel{(3)\quad}{{Y}_{N,j}} + d_8 \ddot{Y}_{K,j}
+d_9 \ddot{Y}_{{K{\rm ae}},j} \bigr )
\nonumber \\
& \quad
+ \epsilon^2  \bigl ( d_{10} \stackrel{(3)\quad}{{Y}_{B,j}} + d_{11} \stackrel{(5)\quad}{{Z}_{N,j}} + d_{12} \stackrel{(4)\quad}{{Z}_{K,j}} + 
d_{13} \stackrel{(4)\qquad}{{Z}_{{K{\rm ae}},j}}  \bigr ) \,,
\nonumber \\
\tilde{K}_{\rm ae}^j &= K_{\rm ae}^j + K^j +e_3 \dot{X}_{N,j} + e_4 {X}_{K,j} + e_5 {X}_{K{\rm ae},j}
\nonumber \\
& \quad
+ \epsilon \bigl ( e_6 \dot{X}_{B,j} + e_7 \stackrel{(3)\quad}{{Y}_{N,j}} + e_8 \ddot{Y}_{K,j}
+e_9 \ddot{Y}_{K{\rm ae},j} \bigr )
\nonumber \\
& \quad
+ \epsilon^2  \bigl ( e_{10} \stackrel{(3)}{{Y}}_{B,j} + e_{11} \stackrel{(5)\quad}{{Z}_{N,j}} + e_{12} \stackrel{(4)\quad}{{Z}_{K,j} } +
e_{13} \stackrel{(4)\qquad}{{Z}_{K{\rm ae},j}}  \bigr) \,,
\label{eq:transform0}
\end{align}
where the various superpotentials are defined by Eqs.\ (\ref{eq:super}).

These arbitrary coefficients can then be chosen so as to eliminate all linear terms in Eqs.\ (\ref{eq:linear1}) and (\ref{eq:linear2}) through 2PN order, leaving only 
$(1 - c_{14}/2) \Box N =0$, $\Box B_{jk} = 0$, $\Box K^j =0$, and $\Box^* K_{\rm ae}^j = 0$, where $\Box^* \equiv \nabla^2 - v_T^{-2} \partial_0^2$.  The resulting solution is given by Eq.\ (\ref{eq:transform}).  Should one wish to go to higher PN order, it is straightforward to extend the linear transformation (at the cost of introducing even more exotic superpotentials), to push the offending linear terms to even higher PN orders.

In terms of the new variables, the harmonic gauge conditions become
\begin{align}
  K^j_{,j} &= -\left (1- \frac{c_{14}}{2} \right )\dot{N} - 2c_1 K^j_{{\rm ae},j} 
 \nonumber \\
 & \quad
 + \epsilon c_1 \left ( 1-\frac{1}{v_T^2} \right ) \left ( \ddot{X}_{K{\rm ae}} + \frac{1}{12}\epsilon \stackrel{(4)\quad}{{Y}_{K{\rm ae}}} \right ) \,,
\nonumber \\
\dot{K}^j &+ B^{jk}_{,k} = 0 \,.
\end{align}
The first gauge condition can be used to eliminate $K^j_{,j}$ and its various superpotentials from the problem.  By applying the inverse Laplacian to this equation and iterating, we obtain, to the required 2PN order, 
\begin{align}
X_{K} &=  -  (1-\frac{c_{14}}{2})\dot{X}_N -2 c_1 X_{Kae}  +\epsilon \frac{c_1}{6} W_T \ddot{Y}_{Kae}\,,
\nonumber \\
Y_{K} &=  -  (1-\frac{c_{14}}{2})\dot{Y}_N -2 c_1 Y_{Kae} \,.
\end{align}
These relations have been used to eliminate $K^j_{,j}$, $X_{K}$ and $Y_{K}$ from the transformations shown in Eq.\ (\ref{eq:transform}). 

\section{Properties of Poisson Potentials}
\label{poisson}

Here we summarize some useful properties of Poisson potentials and
superpotentials, defined in Sec.\ \ref{sec:defpotentials}.  These rely upon the
general result, which can be obtained by integration by parts,
\begin{equation}
P(\nabla^2 g) = -g + {\cal B}_P(g) \,,
\end{equation}
where ${\cal B}_P(g)$ denotes the boundary term, given by
\begin{align}
{\cal B}_P(g) &\equiv \frac{1}{{4\pi}} \oint_{\partial {\cal M}} \biggl
[
\frac{{g(t,{\bf x}^\prime)}}{{|{\bf x}-{\bf x}^\prime |}} \partial_r^\prime
\ln (g(t,{\bf x}^\prime)|{\bf x}-{\bf x}^\prime |) \biggr ]_{r^\prime
={\cal R}}
\nonumber \\
&\qquad \times {\cal R}^2 d\Omega^\prime \,.
\label{Pformulae1}
\end{align}
The boundary terms must be carefully evaluated case by case to
determine if any $\cal R$-independent terms survive.  All $\cal
R$-{\it dependent} terms can be discarded.  At 2.5PN order, none of these surface terms contribute.  Some useful formulae that
result from this include:
\begin{align}
P(|\nabla g|^2) &= -\frac{1}{2} \{ g^2 + 2 P(g\nabla^2 g)  \} \,, 
\nonumber\\
P(\nabla g \cdot \nabla f) &= -\frac{1}{2} \{ fg +  P(f\nabla^2 g) +
 P(g\nabla^2 f)  \} \,,
 \nonumber \\
P(U) &= -\frac{1}{2}X  \,, 
\quad
P(X) = - \frac{1}{12}Y \,, 
\nonumber\\
P(|\nabla U|^2) &= - \frac{1}{2}U^2 + \Phi_2 \,,
\nonumber\\
P(\nabla U \cdot \nabla {\ddot X}) &= -\frac{1}{2} \{ U {\ddot X} -
\Sigma(\ddot X) + 2G_2  \}   \,. 
\end{align}
Other useful identities include

\begin{align}
\Sigma(x^i) &= x^i U -X^{,i} \,, 
\nonumber \\
P(1) &= -\frac{1}{6} r^2 \,,
\nonumber \\
P(x^k) &= -\frac{1}{10} x^k r^2 \,.
\end{align}


\end{document}